\newif\ifconfver
\confvertrue        

\newif\ifonecoltab
\onecoltabtrue        

\newif\ifplainver  

\ifplainver
\confverfalse  
\fi

\ifconfver
\documentclass[10pt,journal]{IEEEtran}
\else
\ifplainver
\documentclass[11pt]{article}
\usepackage{fullpage}
\else
\documentclass[12pt,draftcls,onecolumn]{IEEEtran}
\fi
\fi

\usepackage{calc,amsfonts,amssymb,amsmath,bm,url,color,theorem,graphicx,cite,epstopdf,nicefrac,stmaryrd}
\usepackage{psfrag,subfigure,float,epstopdf}
\usepackage[algoruled,linesnumbered]{algorithm2e}
\usepackage{multirow,booktabs}
\usepackage{algorithmic,bbold}
\usepackage[normalem]{ulem}

\definecolor{orange}{RGB}{255,107,0}

\def\black{\color{black}}

\newtheorem{Fact}{Fact}
\newtheorem{Lemma}{Lemma}
\newtheorem{Prop}{Proposition}

\theoremstyle{definition}

\newtheorem{Assumption}{Assumption}
\newtheorem{Remark}{Remark}

\newcommand{\G}{\boldsymbol{G}}

\newcommand{\X}{\boldsymbol{X}}

\renewcommand{\H}{\boldsymbol{H}}

\newcommand{\A}{\boldsymbol{A}}

\newcommand{\x}{\boldsymbol{x}}

\newcommand{\T}{{\!\top\!}}

\newcommand{\tX}{\underline{\bm X}}

\DeclareMathOperator*{\minimize}{\textrm{minimize}}

\begin{document}
	\newcommand{\papertitle}{
		Block-Randomized Stochastic Proximal Gradient for Low-Rank Tensor Factorization
	}
	
	\newcommand{\paperabstract}{%
		This work considers the problem of computing the \textit{canonical polyadic decomposition} (CPD) of large tensors.
		Prior works leverage data sparsity to handle this problem, which is not suitable for handling dense tensors that often arise in applications such as medical imaging, computer vision, and remote sensing.
		Stochastic optimization is known for its low memory cost and per-iteration complexity when handling dense data.
		However, existing stochastic CPD algorithms are not flexible to incorporate a variety of constraints/regularization terms that are of interest in signal and data analytics. Convergence properties of many such algorithms are also unclear.
		In this work, we propose a stochastic optimization framework for large-scale CPD with constraints/regularization terms.
		The framework works under a doubly randomized fashion, and can be regarded as a judicious combination of \textit{randomized block coordinate descent} (BCD) and \textit{stochastic proximal gradient} (SPG).
		The algorithm enjoys lightweight updates and small memory footprint.
		This framework entails considerable flexibility---many frequently used regularizers and constraints can be readily handled. The approach is supported by convergence analysis. Numerical results on large-scale dense tensors are presented to showcase the effectiveness of the proposed approach.
	}

	\ifplainver
	
	\date{\today}
	
	\title{\papertitle}
	
	\author{
		Xiao Fu$^\ast$, Cheng Cao$^\ast$, Hoi-To Wai$^\dag$, and Kejun Huang$^\star$
		\\ ~ \\
		$^\ast$School of Electrical Engineering and Computer Science\\ Oregon State University\\
		$^\dag$Department of Systems Engineering and Engineering Management\\ The Chinese University of Hong Kong\\
		$^\star$Department of Computer and Information Science and Engineering\\ University of Florida\\
		\\~
	}
	
	\maketitle

	\begin{abstract}
		\paperabstract
	\end{abstract}
	
	\else
	\title{\papertitle}
	
	\ifconfver \else {\linespread{1.1} \rm \fi
		
		\author{Xiao Fu, \IEEEmembership{Member, IEEE}, Shahana Ibrahim, \IEEEmembership{Student Member, IEEE}, Hoi-To Wai, \IEEEmembership{Member, IEEE},\\ Cheng Gao, and Kejun Huang, \IEEEmembership{Member, IEEE}
			
			\thanks{

				This work is supported in part by National Science Foundation under Projects NSF ECCS-1608961, ECCS-1808159, III-1910118, the Army Research Office (ARO) under Proejct ARO W911NF-19-1-0247, and by the Chinese University of Hong Kong under the CUHK Direct Grant 4055113.
				
				X. Fu and S. Ibrahim are with the School of Electrical Engineering and Computer Science, Oregon State University, Corvallis, OR 97331, United States. email (xiao.fu,ibrahish)@oregonstate.edu

				H.-T. Wai is with the Department of Systems Engineering and Engineering Management, Shatin, Hong Kong. email: htwai@se.cuhk.edu.hk

				C. Gao was with the School of Electrical Engineering and Computer Science, Oregon State University, Corvallis, OR 97331, United States. He is now with the University of Missouri - Columbia, Columbia, MO 65211. email: gaoche@oregonstate.edu 
				
				K. Huang is with the Department of Computer and Information Science and Engineering, Gainesville, FL 32611, University of Florida. email: kejun.huang@ufl.edu

			}
		}
		
		\maketitle
		
		\ifconfver \else
		\begin{center} \vspace*{-2\baselineskip}
		\end{center}
		\fi
		
		\begin{abstract}
			\paperabstract
		\end{abstract}
		
		\begin{IEEEkeywords}\vspace{-0.0cm}
			Large-scale tensor decomposition, canonical polyadic decomposition, stochastic gradient, Adagrad
		\end{IEEEkeywords}
		
		\ifconfver \else \IEEEpeerreviewmaketitle} \fi
	
	\fi
	
	\ifconfver \else
	\ifplainver \else
	\newpage
	\fi \fi
	\section{Introduction}\label{sec:intro}
	{\it Canonical polyadic decomposition} (CPD) [previously known as parallel factor analysis (PARAFAC)] \cite{kolda2009tensor,sidiropoulos2017tensor,carroll1970analysis} is arguably the most popular low-rank tensor decomposition model.
	CPD has found applications in various fields, such as analytical chemistry \cite{Rao1999}, social network mining \cite{sun2006beyond}, hyperspectral imaging \cite{kanatsoulis2018hyperspectral}, topic modeling \cite{anandkumar2014tensor}, and time series analysis \cite{anandkumar2012method}; also see \cite{SidBroGia00,Sidiropoulos2001,fu2015factor} for more applications in communications. 
	
	Computing the CPD of a tensor, however, is a challenging optimization problem \cite{hillar2013most}. Many algorithms have been proposed through the years \cite{huang2016flexible,xu2013block,liavas2015parallel,carroll1970analysis}.
	To keep pace with the ever growing volume of available data, one pressing challenge is to compute CPD at scale.
	The classic alternating least squares (ALS) algorithm \cite{carroll1970analysis} has an elegant algorithmic structure, but suffers from a number of numerical issues \cite{navasca2008swamp,comon2009tensor} and is hardly scalable.
	In recent years, many new CPD algorithms have appeared, triggered by the advances in big data analytics and first-order optimization \cite{kang2012gigatensor,papalexakis2012parcube,xu2013block,huang2016flexible,mitchell2018nesterov}. Many of these algorithms leverage data sparsity to scale up CPD---by cleverly using the zero elements in huge tensors, computationally costly key operations in ALS (e.g., the \textit{matricized tensor times Khatri-Rao product} (MTTKRP) operation) can be significantly simplified.
	Consequently, the classic ALS algorithm can be modified to handle CPD of huge and sparse tensors.
	
	However, when the tensor to be factored is {\it dense}---i.e., when most entries of the tensor are nonzero---the sparsity-enabled efficient algorithms \cite{huang2016flexible,xu2013block,kang2012gigatensor,papalexakis2012parcube,kanatsoulis2018hyperspectral} are no longer applicable.
	Note that large and dense tensors arise in many timely and important applications such as medical imaging \cite{alexander2007diffusion}, hyperspectral imaging \cite{kanatsoulis2018hyperspectral}, and computer vision \cite{shashua2005non}.
	In fact, since big dense tensors typically cost a lot of memory (e.g., a dense tensor with a size of $2,000\times 2,000 \times 2,000$ occupies 57.52GB memory if saved as double-precision numbers), it is even hard to load them into the RAM of laptops, desktops, or servers. 

Stochastic gradient (SG) method is a powerful tool for handling optimization problems involving a large amount of data, which is known for its low per-iteration memory and computational complexities \cite{bottou2018optimization}.
	A number of stochastic optimization based CPD algorithms have been proposed in the literature \cite{vervliet2016randomized,beutel2014flexifact,battaglino2018practical}.
	Specifically, The works in \cite{vervliet2016randomized,beutel2014flexifact} work in an iterative manner. 
	In each iteration, the algorithm samples a random subset of the tensor entries and update the corresponding parts of the latent factors using the sampled data.
	The algorithms have proven quite effective in practice, and feature distributed implementation \cite{beutel2014flexifact}.
	The challenge here is that every tensor entry only contains information of a certain \textit{row} of the latent factors, and updating the entire latent factors may require a high complexity.
	This may lead to slow improvement of the latent factor estimation accuracy.
	More importantly, this update strategy loses the opportunity to incorporate constraints/regularization terms on the whole latent factors, since the sampled entries only contain partial information of them.
	This is undesired in practice, since prior information on the latent factors are critical for enhancing performance, especially in noisy cases.

	Recently, a stochastic algorithm that ensures updating one entire latent factor in every iteration was proposed in \cite{battaglino2018practical}. 
	Instead of sampling tensor entries, the algorithm works via sampling {\it tensor fibers} that contain information of the whole latent factors.
	{\black This algorithm exhibits very good empirical performance when the tensor rank is low}.
	However, this algorithm works with at least as many fibers as the tensor rank, which in some cases gives rise to high per-iteration complexity. 
	In addition, the algorithm in \cite{battaglino2018practical} {\black did not explicitly offer implementations that take into considerations of constraints or regularization terms---although this can be fixed with some modifications.}
	Lastly, convergence properties of many stochastic CPD algorithms such as those in \cite{vervliet2016randomized,battaglino2018practical} are unclear.
	
	\noindent
	{\bf Contributions} 
	In this work, we propose a new stochastic algorithmic framework for computing the CPD of large-scale dense tensors. Our contributions include:
	
	\noindent $\bullet$ {\bf A Doubly Randomized Computational Framework for Large-Scale CPD}. We propose an efficient and flexible computational framework for CPD of large dense tensors. Our method is a judicious combination of randomized block coordinate descent (BCD) \cite{beck2013convergence,nesterov2012efficiency} and stochastic proximal gradient (SPG) \cite{ghadimi2016accelerated,ghadimi2013stochastic}. In each iteration, our algorithm first samples a mode from all modes of the tensor.
	Then, the algorithm samples some fibers of this mode and updates the corresponding latent factor via stochastic proximal operations.
	Such a combination exhibits an array of attractive features:
	It admits smaller per-iteration memory and computational complexities relative to the existing fiber sampling based method in \cite{battaglino2018practical}, {\black particularly in high-rank cases}.
	More importantly, it is very flexible in terms of incorporating {\black regularization} terms and constraints on the latent factors.

	\noindent $\bullet$ {\bf Convergence Analysis}. Both BCD and SPG are well studied topics in the optimization literature \cite{razaviyayn2013unified,nesterov2012efficiency,beck2013convergence}. However, convergence properties of the proposed framework is not immediately clear due to the nonconvex nature of CPD. The existing block-randomized SGD (BR-SGD) framework \cite{wang2014randomized} only considers convex optimization.
	Another work \cite{xu2015block} considers nonconvex optimization, which adopts a Gauss-Seidel type block updating strategy without randomization. The conditions for convergence in \cite{xu2015block} are not easy to check or guarantee in the context of tensor factorization. {\color{black}In contrast, we offer tailored convergence analyses of the proposed algorithm leveraging block randomization, and show that the proposed optimization strategy features sub-sequence convergence to a stationary point---which is a \textit{necessary condition} for attaining local or global optima.}

\noindent $\bullet$ {\bf Implementation-friendly Adaptive Stepsize Scheduling}.
One of the most challenging aspects in stochastic optimization is selecting a proper stepsize schedule.
	To make the proposed algorithms friendly to use by practitioners, we propose a practical and adaptive stepsize schedule based on the celebrated \texttt{Adagrad} algorithm \cite{duchi2011adaptive}. \texttt{Adagrad} is an adaptive stepsize selection method devised for single-block gradient descent. Nonetheless, we find through extensive simulations that it largely helps reducing the agonizing pain of tuning stepsize when implementing our multi-block algorithm for CPD. In addition, we show that the adaptive stepsize-based algorithm converges to a stationary point almost surely.
	
\begin{figure}
		\centering
		\includegraphics[width=0.7\linewidth]{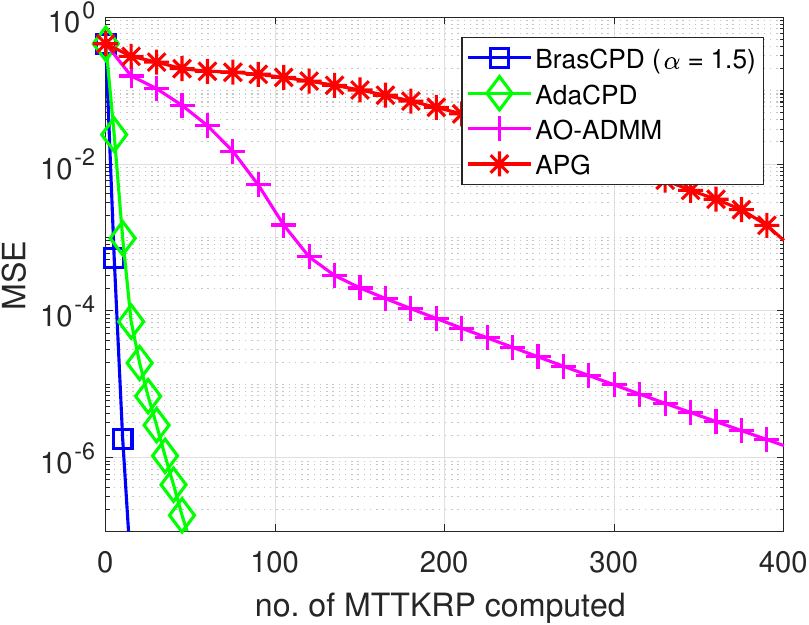}
		\caption{The proposed algorithms (\texttt{AdaCPD} and \texttt{BrasCPD}) exhibit low complexity for achieving high accuracy of the estimated latent factors. The tensor under test has a size of $100\times 100\times 100$ and the rank is 10. The latent factors are constrained to be nonnegative. The baselines are two state-of-art constrained CPD algorithms \texttt{AO-ADMM} \cite{huang2016flexible} and \texttt{APG} \cite{xu2013block}.}
		\label{fig:10010mse}
		\vspace{-.5cm}
	\end{figure}

	A quick demonstration of the effectiveness of the proposed algorithms is shown in Fig.~\ref{fig:10010mse}, where the average mean squared error (MSE) of the estimated latent factors [cf. Eq.~\eqref{eq:mse}] against the number of MTTKRP computed (which dominates the complexity) is plotted. One can see that the proposed algorithm largely outperforms a couple of state-of--the-art algorithms for constrained CPD.
	More thorough numerical results can be seen in Sec.~\ref{sec:num}.
	
	Part of the work was submitted to ICASSP 2019 \cite{fu2018icassp}. In this new version, we have included detailed convergence proofs and the new adaptive stepsize based algorithm. More extensive simulations and real-data experiments are also included.

	\noindent
	{\bf Notation.}
	We follow the established conventions in signal processing. 
	$x$, $\bm x$, $\bm X$, and $\tX$ denote scalar, vector, matrix, and tensor, respectively; $\|\cdot\|$ denotes the Euclidean norm, i.e., $\|\x\|_2$ and $\|\X\|_F$, respectively; $\circ$, $\odot$, and $\circledast$ denote outer product, Khatri-Rao product, and Hadamard product, respectively, unless otherwise specified; ${\rm vec}(\X)$ denotes the vectorization operator that concatenates the columns of $\X$; $\X\geq\bm 0$ means that all the entries of $\X$ are nonnegative; $^\T$ and $^\dag$ denote transpose and pseudo-inverse, respectively; $|{\cal C}|$ denotes the cardinality of set ${\cal C}$; $\lambda_{\max}(\cdot)$ denotes the largest eigenvalue of a matrix. Unless otherwise specified, we denote the total expectation by the subscript-less operator $\mathbb{E}[\cdot]$.

	\section{Background}
	We first introduce some notions used in tensor algebra.
	\subsection{Tensors and CPD}
	An $N$th order tensor is an array whose entries are indexed by $N$ coordinates; i.e., $\underline{\bm X}(i_1,\ldots,i_N)$ denotes an element of the tensor $\underline{\X}$ with a size of $I_1\times I_2\times \ldots \times I_N$. Like matrices, tensors can be represented as sum of rank-one components:
	\begin{equation}\label{eq:cpd}
	\underline{\X} = \sum_{f=1}^F \A_{(1)}(:,f)\circ \A_{(2)}(:,f) \circ \ldots \circ \A_{(N)}(:,f),
	\end{equation}
	where ``$\circ$'' denotes the outer product of vectors, and $\A_{(n)}$ is an $I_n \times F$ matrix that is often referred to as the \textit{mode-$n$ latent factor}.
	When $F$ is the minimal integer that satisfies the expression in \eqref{eq:cpd}, the right hand side in \eqref{eq:cpd} is called the {\it canonical} polyadic decomposition of the tensor $\underline{\X}$. 
	At the entry level, the CPD can be expressed as
	\begin{equation}\label{eq:er}
	\tX(i_1,\ldots,i_N)=\sum_{f=1}^F \prod_{n=1}^N \A_{(n)}(i_n,f),
	\end{equation}
	for $i_n\in\{1,\ldots,I_n\}$. 
	The CPD of a tensor is essentially unique under mild conditions\footnote{The latent factors $\A_{(n)}$'s that constitute the data $\underline{\X}$ are unique up to some trivial ambiguities like column permutations and scalings \cite{sidiropoulos2017tensor}.}. 
The CPD of a tensor can be obtained via minimizing a certain criterion as follows:
	\begin{equation}\label{eq:lscri}
	\minimize_{\{\A_{(n)}\}_{n=1}^N}~f(\A_{(1)},\ldots,\A_{(N)}).
	\end{equation}
A common optimization criterion for CPD in the literature is the \textit{least squares} (LS) fitting criterion \cite{carroll1970analysis,huang2016flexible,xu2013block}:
	\[ f(\A_{(1)},\ldots,\A_{(N)})   = \Big\| \underline{\X} - {\sum_{f=1}^F} \A_{(1)}(:,f)\circ \ldots \circ \A_{(N)}(:,f) \Big\|_F^2.    \]
	In the sequel, we will often use the shorthand notation $f(\bm \theta)$ to denote $f(\A_{(1)},\ldots,\A_{(N)}) $,
	where
	\[ \bm \theta =[{\rm vec}(\A_{(1)})^\T,\ldots,{\rm vec}(\A_{(N)}^\T)]^\T.  \]
Other criteria have also been considered, e.g., the Kullback-Leibler (KL) divergence \cite{chi2012tensors} and robust fitting \cite{RobustParafac,fu2015joint} criteria---which serve for different purposes. 
	\subsection{Unfolding, ALS and MTTKRP}
	The matricization operation, or \textit{matrix unfolding} of a tensor, has proven very useful in designing tensor factorization algorithms. 
	The mode-$n$ unfolding of a tensor is a $J_n \times I_n$ matrix where 
	\[\underline{\X}(i_1,\ldots,i_N)=\X_{(n)}(j,i_n),\] 
	and we have
	$j=1+\sum_{k=1,k\neq n}^N(i_k-1)J_k$ and $J_k =\prod_{m=1,m\neq n}^{k-1}I_m$  \cite{kolda2009tensor}. The CPD representation in Eq.~\eqref{eq:cpd} can be expressed as
	\begin{equation}
	\X_{(n)} = \H_{(n)} \A_{(n)}^\T,
	\end{equation}
	where the $J_n \times F$ matrix $\H_{(n)}$ is defined as
	\[\H_{(n)} =\A_{(1)}\odot \A_{(n-1)}\odot\A_{(n+1)} \odot\ldots \odot \A_{(N)}=\odot_{i=1,i\neq n}^N\A_{(i)}.\] 
	
	The elegant form of the unfoldings has enabled the famous {\it alternating least squares} (ALS) algorithm \cite{carroll1970analysis} for handling Problem~\eqref{eq:lscri} with the LS objective. Specifically, ALS solves the following cyclically for $n=1,\ldots,N$:
	\begin{align}\label{eq:als}
	\A_{(n)} & \leftarrow \arg\min_{\A}~\big\| \X_{(n)} -  \H_{(n)} \A^\T \big\|_F^2.
	\end{align}
	Problem~\eqref{eq:als} is nothing but a least squares problem that admits the following closed-form solution:
	\[  \A_{(n)} \leftarrow \left(   (\H_{(n)}^\T\H_{(n)})^{-1}\H_{(n)}^\T \X_{(n)} \right)^\T, \]
	if ${\rm rank}(\H_{(n)})=F$.
	Note that $ (\H_{(n)}^\T\H_{(n)})^{-1}$ is not difficult to compute by exploiting the Khatri-Rao structure of $\H_{(n)}$ \cite{huang2016flexible,sidiropoulos2017tensor,kolda2009tensor}.
	However, when the problem dimension is large (which often happens in applications such as medical imaging, remote sensing, and computer vision),
	solving the seemingly simple problem in \eqref{eq:als} can be computationally prohibitive. The reason is that both $\bm X_{(n)}\in\mathbb{R}^{(\prod_{j=1,j\neq n}^NI_j )\times I_n}$ and $\H_{(n)}\in\mathbb{R}^{(\prod_{j=1,j\neq n}^NI_j )\times F}$ can be very large matrices.
	In particular, the so-called \textit{matricized tensor times Khatri-Rao product} (MTTKRP) operation, i.e., \[{\sf MTTKRP:}\quad\quad\H_{(n)}^\T \X_{(n)}\] that happens in every iteration of ALS costs ${\cal O}(\prod_{n=1}^NI_nF)$ flops (or, ${\cal O}(I^NF)$ if $I_n=I$). 
	This is quite costly even if $I_n$ is moderately large.
	Many works have considered fast algorithms for computing MTTKRP, but these methods are mainly enabled by judiciously exploiting sparsity of the tensor data \cite{Bulletin2013,kang2012gigatensor}. Computing MTTKRP for dense tensors has also been considered. Nonetheless, these works are often concerned with practical implementation schemes such as parallelization and memory-efficient computation strategies, but the number of computational flops required is naturally high for the dense tensor case; see, e.g., \cite{ravindran2014memory,li2015input}.

	In a lot of applications, some prior knowledge on the latent factors is known---e.g., in image processing, $\A_{(n)}$'s are normally assumed to be nonnegative \cite{kanatsoulis2018hyperspectral}; in statistical machine learning, sometimes the columns of $\A_{(n)}$ are assumed to be constrained within the probability simplex \cite{chi2012tensors,kargas2017tensors}; i.e.,
	\begin{equation}\label{eq:simplex}
	\bm 1^\T\A_{(n)}=\bm 1^\T,~\A_{(n)}\geq \bm 0.
	\end{equation}
	In those cases, the following criterion is often of interest:
	\begin{equation}\label{eq:lsreg}
	\begin{aligned}
	 \minimize_{\{\A_{(n)}\}_{n=1}^N} &~ f(\A_{(1)},\ldots,\A_{(N)}) \color{black}\\
	{\rm subject~to}&~\A_{(n)}\in{\cal A}_n.
	\end{aligned}
	\end{equation}
	Compared to the unconstrained version, Problem~\eqref{eq:lsreg} is even harder to handle.
	Some recent methods combine first-order constrained optimization and ALS \cite{xu2013block,huang2016flexible} to make the tensor factorization algorithms more flexible in handling constraints and regularization terms---but the complexity orders of those algorithms often scale similarly as that of ALS,
	since these algorithms do not avoid computing $\H_{(n)}^\T \X_{(n)}$ that is the bottleneck for computing CPD.
	
	\subsection{Stochastic Optimization} 
	
	When the tensor is large and dense, working with the entire dataset could be computationally and memory-wise expensive. A popular workaround is to apply \textit{stochastic optimization}---i.e., sampling parts of the data at random and use the sampled piece to update the latent factors.
	Using Eq.~\eqref{eq:er}, Problem~\eqref{eq:lscri} with the LS objective is equivalent to the following:
	\begin{equation}\label{eq:sgd_cri}
	\minimize_{\{\A_{(n)}\}}~(1/T) ~{\textstyle \sum_{i_1,\ldots,i_N}}~f_{i_1,\ldots,i_N}\left(\bm \theta\right),
	\end{equation}
	where $T=\prod_{n=1}^NI_n$ and \[f_{i_1,\ldots,i_N}(\bm \theta)=\left( \tX(i_1,\ldots,i_N)-\sum_{f=1}^F \prod_{n=1}^N \A_{(n)}(i_n,f) \right)^2.\]
	The objective function in \eqref{eq:sgd_cri} can be understood as an empirical risk  \cite{bottou2018optimization}.
	Using this observation, the algorithms in \cite{vervliet2016randomized,beutel2014flexifact} randomly sample a {\black subset} of entries indexed by $\{(i_1,\ldots,i_N)\}$ and update the pertinent parts of the latent factors (note that the $(i_1,\ldots,i_N)$th entry of tensor contains the information of $\A_{(n)}(i_n,:)$ for $n=1,\ldots,N$) using the sampled entries of the tensor. For example, \cite{beutel2014flexifact} uses a stochastic gradient (SG) based approach and update the $\A_{(n)}(i_n,:)$'s that are associated with the sampled entries. 
	The sampling method in \cite{vervliet2016randomized} is similar, while the update is not gradient-based but Gauss-Newton or ALS applied to the sampled set of entries (or, sub-tensors, to be precise).
	The upshot of this line of work is that the per-iteration complexity can be quite low. 
	
	Despite of such favorable complexity savings, the approaches in \cite{vervliet2016randomized,beutel2014flexifact} have a couple of limitations.  One challenge is that many useful prior information cannot be incorporated in the algorithm. The reason is that these algorithms update
		part of the {\it rows} of $\A_{(n)}$'s, while many useful priors are defined w.r.t. the \textit{columns} of the latent factors, e.g., the probability simplex constraint in \eqref{eq:simplex} and the total variation constraint that is heavily used in image processing. For example, the algorithm in \cite{vervliet2016randomized} samples subtensors $\tX_{\sf sub}=\tX({\cal S}_1,\ldots,{\cal S}_N)$ (where ${\cal S}_n\subset\{1,\ldots,I_n\}$) to update the corresponding $\A_{(n)}({\cal S}_n,:)$'s. Under such a scheme, it is hard to handle constraints like $\bm 1^\T\A_{(n)}=\bm 1^\T, \A_{(n)}\geq\bm 0$ that are critical in statistical learning \cite{kargas2017tensors,ibrahim2019stochastic,kargas2019learning,traganitis2018blind}.
	Third, convergence properties of these methods are often unclear.
	
\begin{figure}
		\centering
		\includegraphics[width=0.65\linewidth]{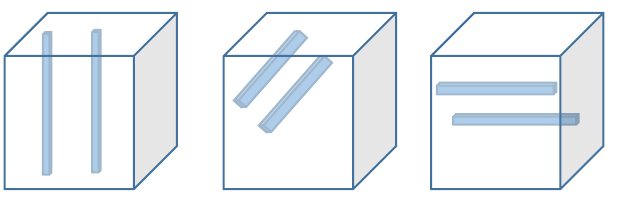}
		\caption{From left to right: mode-$1,2,3$ tensor fibers of a third-order tensor, respectively.}\vspace{-.2cm}
		\label{fig:fibers}
		\vspace{-.2cm}
	\end{figure}

	An alternative \cite{battaglino2018practical} is to leverage the tensor data structure by considering 
	randomly sampled \emph{fibers} of tensors. 
	Note that a mode-$n$ fiber of $\underline{\X}$ (cf.~Fig.~\ref{fig:fibers}) is a row of the mode-$n$ unfolding 
	$\X_{(n)}$ \cite{kolda2009tensor}. 
	Now, 
	assuming that one samples a set of mode-$n$ fibers indexed by ${\cal F}_n \subset \{1,...,J_n\}$, then $\A_{(n)}$ can be updated by solving a `sketched version' of Problem~\eqref{eq:als}:
	\begin{align}\label{eq:sktech_als}
	\A_{(n)} & \leftarrow \arg\min_{\A}~\big\| \X_{(n)}({\cal F}_n,:) -  \H_{(n)}({\cal F}_n,:) \A^\T \big\|_F^2,
	\end{align}
	If $|{\cal F}_n|\geq F$, then the sketched system of linear equations $ \X_{(n)}({\cal F}_n,:) \approx \H_{(n)}({\cal F}_n,:) \A_{(n)}^\T $ {\black can be} over-determined.
	Hence, one can update $\A_{(n)}$ by solving the $|{\cal F}_n|$ dimensional linear system 
	\begin{equation}\label{eq:cprandls}
	\A_{(n)}^\T \leftarrow  \H_{(n)}({\cal F}_n,:)^\dag \X_{(n)}({\cal F}_n,:).
	\end{equation}
	Similar to the ALS algorithm, after updating $\A_{(n)}$, the algorithm moves to mode-$(n+1)$ fibers and repeats the same for updating $\A_{(n+1)}$.
    The downside of this method is that it needs to sample at least $F$ fibers for each update, and $F$ can be larger than $I_n$ in tensor decomposition.
	{\black In addition, the work in \cite{battaglino2018practical} focused on unconstrained cases while did not offer implementations for constrained/regularized cases. This can be compensated by replacing \eqref{eq:cprandls} with a constrained least squares solver or certain constraint enforcing operations.
	However, convergence properties of doing so are also unclear.}

	\section{Proposed Algorithm}
	
	In this work, we propose a new stochastic optimization strategy for CPD. Our method combines the insights from ALS and fiber sampling, but allows $|{\cal F}_n|\ll F$. This is instrumental in practice, since it is the key for achieving low per-iteration complexity.
	The proposed algorithm can easily handle a variety of constraints and regularizations that are commonly used in signal processing and data analytics---which is reminiscent of\textit{ stochastic proximal gradient} (SPG) \cite{ghadimi2013stochastic,xiao2014proximal}.
	In addition, we provide convergence analyses to back up the proposed approach.
	
	\vspace{-.25cm}
	\subsection{Basic Idea: Unconstrained Case}
	
	We first consider Problem \eqref{eq:lscri}. Our idea is to apply SA while exploiting the tensor fiber structure. 
	Specifically, at each iteration, we sample a set of mode-$n$ fibers for a certain $n$ as the method in \cite{battaglino2018practical} does.
	However, instead of exactly solving the least squares subproblems \eqref{eq:als} for all the modes following a Gauss-Seidel manner in each iteration, we update $\A_{(n)}$ using a doubly stochastic procedure. To be more precise, at iteration $r$, we first randomly sample a mode index $n \in \{1,...,N\}$.
	Then, we randomly sample a set of mode-$n$ fibers that is indexed by ${\cal F}_{n} \subset
	\{1,...,J_n \}$.
	Let ${\bm G}^{(r)} \in \mathbb{R}^{(I_1+\ldots+I_N) \times F} $ such that
		$$ {\bm G}^{(r)} = [(\G_{(1)}^{(r)})^\T,\ldots,(\G_{(N)}^{(r)})^\T]^\T, $$
		where we have
		\begin{align} 
		&\notag {\bm G}_{(n)}^{(r)} = \frac{1}{|{\cal F}_n|} \Big( {\bm A}_{(n)}^{(r)} {\bm H}_{(n)}^\T({{\cal F}_n}){\bm H}_{(n)}({\cal F}_n) - {\bm X}_{(n)}^\T ({\cal F}_n) {\bm H}_{(n)}({\cal F}_n) \Big)\\
		& {\bm G}_{(n')}^{(r)} = {\bm 0} ,~~n' \neq n, \label{eq:als_sgd}
		\end{align}
		and we used the shorthand notations $${\bm X}_{(n)} ({\cal F}_n) =
		{\bm X}_{(n)} ({\cal F}_n,:),\quad {\bm H}_{(n)}({{\cal F}_n}) = {\bm H}_{(n)}({{\cal F}_n},:).$$
	The latent variables are updated by 
	\begin{equation}\label{eq:proposed}
	\begin{aligned}
	{\bm A}_{(n)}^{(r+1)} \leftarrow \A_{(n)}^{(r)} - \alpha^{(r)}{\bm G}_{(n)}^{(r)},~~n=1,...,N.
	\end{aligned}
	\end{equation}
Observe that ${\bm G}_{(n)}^{(r)}$ is an estimate of the gradient applied to $f(\A_{(1)},\ldots,\A_{(N)})$ taken w.r.t.~the mode-$n$ variable $\A_{(n)}$, and the update is an iteration of the SG algorithm with a minibatch size $|{\cal F}_n|$ for solving the problem in \eqref{eq:als}.
	
	The proposed update is very efficient, since the most resource-consuming update ${\bm H}_{(n)}^\T{\bm X}_{(n)}$ in algorithms such as those in \cite{huang2016flexible,xu2013block} is avoided. The corresponding part ${\bm X}_{(n)}^\T ({\cal F}_n,:){\bm H}_{(n)}({\cal F}_n,:)$ costs only ${\cal O}(|{\cal F}_n|FI_n)$ flops---and $|{\cal F}_n|$ is under our control. Note that the first step in this procedure is different from standard ALS-type algorithms that update the block variables $\A_{(n)}$ cyclically instead of updating a randomly sampled block. 
	As we will show, this modification greatly simplifies our convergence analysis.

	\subsection{Constrained and Regularized Case}
	As mentioned, there are many cases in practice where considering regularizations or constraints on $\A_{(n)}$'s can benefit the associated tasks.
	Since our framework updates an entire $\A_{(n)}$ in each iteration, it is friendly for incorporating a large variety of commonly used constraints/regularizations---which is more flexible relative to the entry sampling based approaches in \cite{vervliet2016randomized,beutel2014flexifact}. The algorithm can be easily extended to handle the constrained/regularized case. Consider:
	\begin{equation}\label{eq:lscri_reg}
	\begin{aligned}
	\minimize_{\{\A_{(n)}\}_{n=1}^N}& ~f(\bm \theta)+\sum_{n=1}^N h_n (\A_{(n)})
	\end{aligned}
	\end{equation}
	where $h_n (\A_{(n)} )$ denotes a structure-promoting regularizer on $\A_{(n)}$.
	Note that if $\A_{(n)}\in{\cal A}_n$ is desired, we can write $h_n(\cdot)$ is defined  as the indicator function of set ${\cal A}_n$:	
	\[ h_n(\bm A)={\cal I}({\cal A}_n)=\begin{cases}
	0,\quad&\A\in{\cal A}_n\\
	\infty,\quad& {\rm otherwise}.
	\end{cases} \]
	Using the same fiber sampling strategy as in the previous subsection, we update $\A_{(n)}$ by
	\begin{subequations}
		\begin{align}
		{\bm A}_{(n)}^{(r+1)}&\leftarrow \arg\min_{\A_{(n)}}~ \big\|\A_{(n)}-\big( \A_{(n)}^{(r)} - \alpha^{(r)}{\bm G}_{(n)}^{(r)} \big) \big\|_F^2 \nonumber\\
		&\hspace{1.5cm} +h_n\left(\A_{(n)}\right), \label{eq:prox}\\
		{\bm A}_{(n')}^{(r+1)}&\leftarrow{\bm A}_{(n')}^{(r)},\quad n' \neq n.
		\end{align}
	\end{subequations}
	If $h_n(\cdot)$ is {\black a closed proper convex function}, the update~\eqref{eq:prox} can be solved by applying the proximal operator of $h_n(\cdot)$, which is often denoted as
	\begin{equation}\label{eq:prox_sgd}
	\A_{(n)}^{(r+1)} \leftarrow {\sf Prox}_{h_n}\left(  \A_{(n)}^{(r)} - \alpha^{(r)}{\bm G}_{(n)}^{(r)}  \right). 
	\end{equation}   
	Many $h_n(\cdot)$'s  admit simple closed-form solutions for their respective proximal operators, e.g., when $h_n(\cdot)$ is the indicator function of the nonnegative orthant and $h_n(\cdot)=\|\cdot\|_1$; see Table~\ref{tab:prox} and more details in \cite{huang2016flexible,parikh2013proximal}.
	The complexity of computing \eqref{eq:prox_sgd} is similar to that of the plain update in \eqref{eq:proposed}, and thus is also computationally efficient.
	An overview of the proposed algorithm can be found in
	Algorithm~\ref{algo:BR-SGD}, which we {name} {\it \textbf{B}lock-\textbf{Ra}ndomized \textbf{S}GD for \textbf{CPD}} (\texttt{BrasCPD}).
	
	\vspace{-.2cm}
	\begin{algorithm}[!t] 
		
		{\small
			\SetKwInOut{Input}{input}
			\SetKwInOut{Output}{output}
			\SetKwRepeat{Repeat}{repeat}{until}
			
			\Input{$N$-way tensor ${\tX}\in\mathbb{R}^{I_1\times \ldots \times I_N}$; rank $F$; sample size $B$, initialization $\{ \A_{(n)}^{(0)} \}$, step size $\{\alpha^{(r)}\}_{r=0,\ldots}$}
			
			$r\leftarrow 0$;
			
			\Repeat{some stopping criterion is reached}{
				uniformly sample $n$ from $\{1,\ldots,N\}$, then 
				sample ${\cal F}_n$ uniformly from $\{1,\ldots, J_n \}$ with $|{\cal F}_n| =B$;
				
				form the stochastic gradient ${\bm G}^{(r)} \leftarrow \eqref{eq:als_sgd}$;
				
				update $\A_{(n)}^{(r+1)} \leftarrow \eqref{eq:prox}$, $\A_{(n')}^{(r+1)} \leftarrow \A_{(n')}^{(r)}$ for $n'\neq n$;
				
				$r\leftarrow r+1$;
			}
			\Output{$\{\A_{(n)}^{(r)}\}_{n=1}^N$}
			
			\caption{\texttt{BrasCPD}}\label{algo:BR-SGD}	\vspace{-.1cm}
		}
		
	\end{algorithm}

	\begin{table}[htbp]
		{	\centering
			
			\caption{Proximal/projection operator of some frequently used regularizations and constraints. }	\label{tab:prox}
			\begin{tabular}{c|c|c}
				\hline
				\hline
				$h(\cdot)$  & prox./proj.  solution & complexity\\
				\hline
				\hline
				$\|\cdot\|_1$    & soft-thresholding  & ${\cal O}(d)$ \\
				\hline
				$\|\cdot\|_2$    &   re-scale    & ${\cal O}(d)$ \\
				\hline
				$\|\cdot\|_{2,1}$ & block soft-thresholding & ${\cal O}(d)$ \\
				\hline
				$\|\cdot\|_0$     & hard-thresholding & ${\cal O}(d)$ \\
				\hline
				${\cal I}(\varDelta)$ &  randomized pivot search \cite{duchi2008efficient}    & ${\cal O}(d)$ in expectation\\
				\hline
				${\cal I}(\mathbb{R}_+)$   & max   & ${\cal O}(d)$ \\
				\hline
				monotonic & monotone regression \cite{kruskal1964nonmetric} & ${\cal O}(d)$ \\
				\hline
				unimodal & unimodal regression \cite{bro1998least} & ${\cal O}(d^2)$ \\
				\hline	\end{tabular}}
		
		\noindent
		$^\dag$In the table, $d$ is the number of optimization variables.
	\end{table}%

	\section{Convergence Properties}
In this section, we offer tailored convergence analyses for \texttt{BrasCPD}.	
To this end, two most relevant works from the optimization literature are \cite{xu2013block} and \cite{wang2014randomized}. 
		The work in \cite{wang2014randomized} considers block-randomized SGD, but only for the convex case---while our problem is nonconvex.
		The work in \cite{xu2013block} considers the Gauss-Seidel type block SGD  (i.e., cyclically updating the blocks), instead of the block-randomized version as \texttt{BrasCPD} uses. There, convergence is established using a number of assumptions that are not easy to check or guarantee in the context of CPD, e.g., that the bias of the stochastic oracle is bounded. We will show that, by using the block-randomization strategy and the proposed stochastic oracle construction, such an assumption can be circumvented\footnote{We should note that the major motivation for using block randomization strategy is theoretical guarantees---since our goal is a convergence-guaranteed algorithmic framework; in practice, we observe cyclically updating the latent factors works as well.}.

	To facilitate our discussions, let us define $\xi^{(r)} \in \{1,...,N\}$ and $\bm{\zeta}^{(r)} \subseteq \{1,...,J_{\xi_{(r)}} \}$ as the random variables (r.v.) responsible for selecting the mode and fibers in iteration $r$, respectively.
		These r.v.s are distributed as
		\begin{equation}
		{\sf Pr}( \xi^{(r)} = n ) = \frac{1}{N} ,~{\sf Pr}( \bm{\zeta}^{(r)} = {\cal S} ~|~ \xi^{(r)} = n ) = \frac{1}{ {J_{n} \choose B} },
		\end{equation}
		where $n \in \{1,...,N\}$, ${\cal S} \in \Sigma$ such that $\Sigma$ is the set of {all}  subsets of $\{1,...,J_{\xi_{(r)}} \}$ with size $B$. We observe
		\begin{Fact}\label{fact:unbias}
			{\color{black} Denote ${\cal B}^{(r)}$ as the filtration generated by the r.v.s $\{ \xi^{(1)},\bm{\zeta}^{(1)}, \ldots,\xi^{(r-1)},\bm{\zeta}^{(r-1)} \}$ such that the $r$th iterate $\bm{\theta}^{(r)}$ is determined conditioned on ${\cal B}^{(r)}$}. The stochastic gradient  in \eqref{eq:als_sgd} is an unbiased estimate for the full gradient w.r.t. $\A_{(\xi^{(r)})}$
			\begin{equation} \label{eq:factunbias}
			\mathbb{E}_{\bm\zeta^{(r)}}\left[  \G_{(\xi^{(r)})}^{(r)}~|~{\cal B}^{(r)},\xi^{(r)} \right] = \nabla_{\A_{(\xi^{(r)})}} f({\bm \theta}^{(r)}).
			\end{equation}
	\end{Fact}
	The proof of the above is straightforward and thus skipped. 
	Fact~\ref{fact:unbias} says that  our block stochastic gradient 
	is an unbiased estimation for the ``block gradient'' $\nabla_{\A_{(n)}} f({\bm \theta}^{(r)})$. This fact will prove quite handy in establishing convergence.
	The two-level sampling strategy (i.e., block sampling and fiber sampling, respectively)  makes the gradient estimation w.r.t. $\bm \theta$ unbiased up to a scaling factor (see Appendix~\ref{app:G}). This connection intuitively suggests that the proposed algorithm should behave similarly as an SG algorithm.
	
\subsection{Unconstrained Case}
We will use the following assumptions:
	\begin{Assumption}\label{as:stepsize}
		The stepsize schedule follows the Robbins-Monro rule \cite{robbins1985stochastic}:
		\[ \textstyle \sum_{r=0}^\infty \alpha^{(r)} =\infty,\quad    \sum_{r=0}^\infty (\alpha^{(r)})^2 <\infty. \]
	\end{Assumption}
	\begin{Assumption}\label{as:boundedness}
		The updates $\A_{(n)}^{(r)}$ are bounded for all $n,r$.
	\end{Assumption}

		Assumption~\ref{as:stepsize} is a principle for stepsize scheduling, which is commonly used in SG algorithms. Assumption~\ref{as:boundedness} is considered a relatively strong assumption. In practice, there are several simple ways to make $\A_{(n)}^{(r)}$'s bounded. One pragmatic modification is to change the objective to $f(\bm \theta)+\sum_{n=1}^Nh_n(\A_{(n)} +\sum_{n=1}^N \lambda_n\|\A_{(n)}\|_F^2$. Another method is as mentioned in \cite{razaviyayn2013unified,huang2016flexible}. At iteration $r$, one may add a proximal term $\lambda_n\|\A_{(n)}-\A_{(n)}^{(r)}\|_F^2$ to the cost function, which will effectively prevent $\A_{(n)}^{(r+1)}$ from being unbounded. Following both ways, the updates can still be handled by simple proximal operations for the $h$ functions in Table~\ref{tab:prox}.

	There are an array of problem structures that are useful for studying convergence of the algorithm.
	\begin{Fact}\label{fact:Lcon} For any $\bm{\theta}, \overline{\bm{\theta}}$ and mode $n \in \{1,...,N\}$, there exists a constant $\bar{L}_{(n)}$ such that 
		\begin{align}\label{eq:lip}
		f(\bm \theta) & \leq f(\bar{\bm \theta}) + \langle \nabla_{\A_{(n)}} f(\bar{\bm \theta}),\A -\bar{\A}_{(n)}\rangle \nonumber \\
		& \quad\quad\quad + \frac{\bar{L}_{(n)}}{2}\|\A -\bar{\A}_{(n)}\|_F^2,
		\end{align}
		where 
		$\bar{\A}_{(n)}$ and $\bar{\H}_{(n)}$ are extracted/constructed from $\bar{\bm \theta}$ following the respective definitions.
	\end{Fact}
	Eq.~\eqref{eq:lip} holds because the objective function $f(\bm \theta)$ w.r.t. $\A_{(n)}$ is a plain least squares fitting criterion, which is known to have a Lipschitz continuous gradient---and the smallest Lipschitz constant is $\lambda_{\max}(\bar{\H}_{(n)}^\T\bar{\H}_{(n)})$.

	We have the following convergence property for \texttt{BrasCPD} in the unconstrained case:
	\begin{Prop}\label{prop:unconstrained}
		Consider the case where $h_n( \cdot ) = 0$ for all $n$ and Assumptions~\ref{as:stepsize}-\ref{as:boundedness} hold. The solution sequence produced by \texttt{BrasCPD} satisfies:
		\[  \displaystyle \liminf_{r \rightarrow \infty} \mathbb{E} \big[  \big\|\nabla f (\bm \theta^{(r)} ) \big\|^2 \big]= 0.\]
	\end{Prop}
The proof is relegated to Appendix~\ref{app:proof_unsconstrained}.
{\color{black}The above proposition implies that there exists a subsequence of the solution sequence that converges to a stationary point in expectation.}
The use of the expectation notion is due to the randomness in the algorithm. We should mention that the SG update and the block sampling step are essential for establishing convergence---and using the exact solution to \eqref{eq:sktech_als} as in \cite{battaglino2018practical} may not have such convergence properties.

	\subsection{Constrained/Regularized Case}

{\color{black} When $h_n(\cdot) \neq 0$, the gradient of the objective function of \eqref{eq:lscri_reg} may be undefined.} {\color{black}In this case, a solution $\bm{\theta}^{(r)}$ is stationary if $\bm P_{(n)}^{(r)} =\bm 0$, $\forall~n$,}
	where 
	$$\bm P_{(n)}^{(r)} = \frac{1}{\alpha^{(r)}} \left(  \A_{(n)}^{(r+1)} - {\sf Prox}_{h_n}\left(  \A_{(n)}^{(r)} - \alpha^{(r)}\nabla_{\A_{(n)}} f(\bm \theta^{(r)})  \right) \right);$$  
	i.e., the condition is satisfied in a blockwise fashion \cite{razaviyayn2013unified,xu2015block}.
	Hence, our goal of this section is to show $\mathbb{E}[\|\bm P_{(n)}^{(r)}\|^2 ]$ vanishes for all $n$ as $r \rightarrow \infty$.
Consider the following assumption:
	\begin{Assumption}\label{as:noisebound}
		There exists a sequence $\{ \sigma^{(r)} \}_{r \geq 0}$ such that
		\[  \mathbb{E}_{\bm\zeta^{(r)}}\left[\left\|\G_{(\xi^{(r)})}^{(r)}-\nabla_{\A_{(\xi^{(r)})}}f(\bm \theta^{(r)})\right\|^2 \left| {\cal B}^{(r)},\xi^{(r)}\right.\right]\leq(\sigma^{(r)})^2, \]
		\begin{equation}\label{eq:noisebound} \textstyle
		\sum_{r=0}^\infty (\sigma^{(r)})^2<\infty, \quad \textstyle
		\sum_{r=0}^\infty \alpha^{(r)}(\sigma^{(r)})^2<\infty,
		\end{equation}
		where $\{\alpha^{(r)}\}_{r\ge 0}$ is the stepsize sequence following Assumption \ref{as:stepsize}.
	\end{Assumption}

The \texttt{BrasCPD} produces a convergent solution sub-sequence:
	\begin{Prop}\label{prop:prox}
		Assume that Assumptions~\ref{as:stepsize}-\ref{as:noisebound} hold. Also assume that $h_n(\cdot)$ is {\black a closed proper convex function}. Then, the solution sequence produced by \texttt{BrasCPD} satisfies
		\[  	  \displaystyle \liminf_{r \rightarrow \infty}\mathbb{E} \Big[ \Big\|\bm P_{(n)}^{(r)} \Big\|^2 \Big]= 0,~\forall ~n. \]
	\end{Prop}
\begin{Remark}
The convergence result in Proposition~\ref{prop:prox} inherits a drawback from single-block stochastic proximal gradient algorithms for nonsmooth nonconvex optimization. To be specific, the relatively strong assumption~\ref{as:noisebound} is required to ensure convergence.
		Assumption~\ref{as:noisebound} implies that the variance of the gradient estimation error $\bm \delta_{(\xi^{(r)})}^{(r)}=\G_{(\xi^{(r)})}^{(r)}-\nabla_{\A_{(\xi^{(r)})}}f(\bm \theta^{(r)})$ converges to zero. This is not entirely trivial. One way to fulfill this assumption is to increase the minibatch size along the iterations, e.g., by setting \cite{ghadimi2013stochastic,xu2015block}:
		\[    |{\cal F}_n^{(r)}| = {\cal O}(\lceil r^{1+\epsilon} \rceil),\quad \forall \epsilon >0.   \]
		Another popular way for achieving \eqref{eq:noisebound} is to use some advanced variance reduction techniques such as SVRG \cite{xiao2014proximal}---which may go beyond the scope of this paper and thus is left out of the discussion.
		Also notice that as the convergence analysis is pessimistic, in practice constant minibatch size works fairly well---as we will see soon.
\end{Remark}

	\section{An Adaptive Stepsize Scheme}
	One may have noticed that the convergence theories in Propositions~\ref{prop:unconstrained}-\ref{prop:prox} do not specify the sequence $\alpha^{(r)}$ except for the two constraints in Assumption~\ref{as:stepsize}. This often gives rise to agonizing tuning experience for practitioners when implementing stochastic algorithms.
	
	Recently, a series of algorithms were proposed in the machine learning community for adaptive stepsize scheduling when training deep neural networks \cite{kingma2014adam,zeiler2012adadelta,dozat2016incorporating}.
	Most of these works are variants of the \texttt{Adagrad} algorithm \cite{duchi2011adaptive}. The insight of \texttt{Adagrad} can be understood as follows: If one optimization variable has been heavily updated before, then it is given a smaller stepsize for the current iteration (and a larger stepsize otherwise). 
	This way, all the optimization variables can be updated in a balanced manner.
	\texttt{Adagrad} was proposed for single-block algorithms, and this simple strategy  admits many provable benefits under the context of convex optimization \cite{duchi2011adaptive}.
	For our multi-block nonconvex problem, we extend the idea and propose the following updating rule:
	In iteration $r$, if $\xi^{(r)}=n$, then, for all $i\in\{1,\ldots,I_n\}$ and all $f\in\{1,\ldots,F\}$, we have
	\begin{subequations}
		\begin{align}
	{\black 	[\bm \eta^{(r)}_{(n)}]_{i,f}} &{\black \leftarrow \frac{\eta}{\left(b + \sum_{t=1}^{r}[{\bm G}_{(n)}^{(t)}]_{i,f}^2\right)^{1/2+\epsilon}},}\label{eq:etaada}\\
	 	\bm A_{(n)}^{(r+1)}& \leftarrow {\sf Prox}_{h_n}\left(\A_{(n)}^{(r)} - \bm \eta^{(r)}_{(n)} \circledast {\bm G}_{(n)}^{(r)}\right),\label{eq:Aada}\\
		\bm A_{(n')}^{(r+1)}&\leftarrow \A_{(n')}^{(r)},~~n' \neq n,
		\end{align}
	\end{subequations}
	where $\eta,b,\epsilon>0$. 
	We note that $b>0$, $\epsilon>0$ are technical conditions used for establishing theoretical convergence. In practice, setting $b=\epsilon=0$ does not hurt the performance and we also observe a slight gain in runtime performance when $\epsilon=0$.
	The \texttt{Adagrad} version of block-randomized CPD algorithm is very simple to implement.
	The algorithm is summarized in Algorithm~\ref{algo:AdaCPD}, which is named \texttt{AdaCPD}.
	
	As one will soon see, such a simple stepsize strategy is very robust to a large number of scenarios under test---i.e., in most of the cases, \texttt{AdaCPD} performs well without tuning the stepsize schedule.
	In addition, the \texttt{AdaCPD} algorithm works well for both the constrained and unconstrained case.
	
	Proving convergence for nonconvex \texttt{Adagrad}-like algorithms is quite challenging \cite{li2018convergence,chen2018convergence}. We show that:
	\begin{Prop}\label{prop:adagrad}
		Assume $h_n(\cdot)=0$ for all $n$, and that ${\sf Pr}(\xi^{(r)}=n) = 1/N$ for all $r$ and $n$.
		Under the Assumptions \ref{as:stepsize}-\ref{as:boundedness}, the solution sequence produced by \texttt{AdaCPD} satisfies
		\[  {\sf Pr}\left( \displaystyle \liminf_{r\rightarrow \infty}\| \nabla f(\bm \theta^{(r)})\|^2 =0  \right)=1. \]
	\end{Prop}
	Proposition~\ref{prop:adagrad} asserts that the algorithm converges almost surely.
	The proof is relegated to Appendix \ref{eq:adacpd}. Our proof extends the idea from a recent paper \cite{li2018convergence} that focuses on using \texttt{Adagrad} for solving single-block nonconvex problems. As mentioned, our two-level sampling strategy makes our algorithm very similar to single-block SGD with a scaled gradient estimation (cf. Appendix~\ref{app:G}), and thus with careful modifications, the key proof techniques in \cite{li2018convergence} goes through. Nevertheless, we detail the proof for being self-containing.

	\vspace{-.2cm}
	\begin{algorithm}[!t] 
		
		{\small
			\SetKwInOut{Input}{input}
			\SetKwInOut{Output}{output}
			\SetKwRepeat{Repeat}{repeat}{until}
			
			\Input{$N$-way tensor ${\tX}\in\mathbb{R}^{I_1\times \ldots \times I_N}$; rank $F$; sample size $B$ , initialization $\{ \A_{(n)}^{(0)} \}$}
			
			$r\leftarrow 0$;
			
			\Repeat{some stopping criterion is reached}{
				uniformly sample $n$ from $\{1,\ldots,N\}$, then 
				sample ${\cal F}_n$ {uniformly} from $\{1,\ldots, J_n \}$ with $|{\cal F}_n| =B$;
				
				form the stochastic gradient ${\bm G}^{(r)} \leftarrow \eqref{eq:als_sgd}$;
				
				determine the step size $\bm \eta^{(r)}_{(n)} \leftarrow \eqref{eq:etaada}$
				
				update $\A_{(n)}^{(r+1)} \leftarrow \eqref{eq:Aada}$, $\A_{(n')}^{(r+1)} \leftarrow \A_{(n')}^{(r)}$ for $n'\neq n$;
				
				$r\leftarrow r+1$;
			}
			\Output{$\{\A_{(n)}^{(r)}\}_{n=1}^N$}
			
			\caption{\texttt{AdaCPD}}\label{algo:AdaCPD}	\vspace{-.1cm}
		}
		
	\end{algorithm}
	
	\color{black}

	\section{Numerical Results}\label{sec:num}
	
	In this section, we use simulations and real-data experiments to showcase the effectiveness of the proposed algorithm.
	
	\subsection{Synthetic Data Simulations}
	\subsubsection{Data Generation}
	Throughout this subsection, we use synthetic third-order tensors (i.e., $N=3$) whose latent factors are drawn from i.i.d. uniform distribution between $0$ and $1$---unless otherwise specified.
	This way, large and {\it dense} tensors can be created.
	For simplicity, we set $I_n=I$ for all $n$ and test the algorithms on tensors having different $I_n$'s and $F$'s.
	In some simulations, we also consider CPD for noisy tensors, i.e., factoring data tensors that have the following signal model:
	\[  \underline{\bm Y}  = \underline{\bm X} + \underline{\bm N},  \]
	where $\underline{\bm X} $ is the noiseless low-rank tensor and  $\underline{\bm N}$ denotes the additive noise. We use zero-mean i.i.d. Gaussian noise with variance $\sigma_N^2$ in our simulations, and the signal-to-noise ratio (SNR) (in dB) is defined as
	$ {\sf SNR} =10\log_{10}\left(\frac{\frac{1}{\prod_{n=1}^NI_n}\|\tX\|^2}{\sigma^2_N} \right)$.

	\subsubsection{Baselines} A number of baseline algorithms are employed as benchmarks.
	Specifically, we mainly use the \texttt{AO-ADMM} algorithm \cite{huangprincipled} and the \texttt{APG} algorithm \cite{xu2013block} as our baselines since they are the most flexible algorithms with the ability of handling many different regularizations and constraints.
	We also present the results output by
	the \texttt{CPRAND} algorithm \cite{battaglino2018practical}. 
	Note that we are preliminarily interested in constrained/regularized CPD. 
	Because \texttt{CPRAND} operates without constraints, the comparison is not entirely fair (e.g., \texttt{CPRAND} can potentially attain smaller cost values since it has a much larger feasible set if other algorithms operate with constraints).
	Nevertheless, we employ it as a benchmark since it uses the same fiber sampling strategy as ours. 
	{\black To make the comparisons more comprehensive, we also offer a simple modification for \texttt{CPRAND} to incorporate constraints/regularization terms. Specifically, we apply the proximal operators associated with the constraint/regularization terms to the original \texttt{CPRAND} updates, and we denote this baseline as \texttt{CPRAND-Prox}.}
	All the algorithms are initialized with the same random initialization; i.e., $\A_{(0)}$'s entries follow the uniform distribution between 0 and 1.
	
	\subsubsection{Parameter Setting}
	For \texttt{BrasCPD}, we set 
	\begin{equation}\label{eq:schedule}
	\alpha^{(r)}=\frac{\alpha}{r^{\beta}},
	\end{equation} where $r$ is the number of iterations, $\beta=10^{-6}$ and $\alpha$ typically takes a value in between 0.001 and 0.1,
	and we try multiple choices of $\alpha$ in our simulations. Our experience is that, under such settings, $\alpha$ is the main tuning parameter that affects the performance of \texttt{BrasCPD}.
	The batch size $|{\cal F}_n|$ is typically set to be below 25  throughout this section.
	For \texttt{AdaCPD}, we fix $b=10^{-6}$, $\epsilon=0$, and $\eta=1$ for all the simulations.
	For \texttt{CPRAND}, we follow the instruction in the original paper \cite{battaglino2018practical} and sample $10F\log_2 F$ fibers for each update.
	
	\subsubsection{Performance Metrics}
	To measure the performance, we employ two metrics.
    We mainly use the estimation accuracy for the latent factors, $\A_{(n)}$ for $n=1,\ldots,N$, as our performance indicator. The accuracy is measured by the \textit{mean squared error} (MSE) which is as defined in \cite{fu2015blind,SOBIUM}:
	\begin{align}
	&{\sf MSE}= \label{eq:mse}\\
	&\min_{\pi(f)\in\{1,\ldots,F\}}\frac{1}{F}\sum_{f=1}^F\left\| \frac{\bm A_{(n)}(:,{\pi(f)})}{\|\bm A_{(n)}(:,{\pi(f)})\|_2} -\frac{\widehat{\A}_{(n)}(:,f)}{\|\widehat{\A}_{(n)}(:,f)\|_2}\right\|_2^2\nonumber
	\end{align}
	where $\widehat{\A}_{(n)}$ denotes the estimate of $\bm A_{(n)}$ and $\pi(f)$'s are under the constraint $\{\pi(1),\ldots,\pi(F)\}=\{1,\ldots,F\}$---which is used to fix the intrinsic column permutation in CPD.  We also use the cost function, i.e.,
	$ {\sf cost}=\left( \nicefrac{1}{\prod_{n=1}^N I_n}\right)\times f(\bm \theta^{(r)})$ as a reference in some of the simulation tables.
	
	Since the algorithms under test have very different operations and subproblem-solving strategies, it may be challenging to find an exactly unified complexity measure.
	In this section, we show the peformance of the algorithms against the number of MTTKRP operations $\H_{(n)}^\T\X_{(n)}$ used, since $\H_{(n)}^\T\X_{(n)}$ is the most costly step that dominates the complexity of all the algorithms under comparison.
	For the stochastic optimization/sketching based algorithms, we also test the MSE/cost value of the algorithms against runtime and the number of sampled entries for updating the latent factors. The latter is particularly meaningful under the stochastic settings, since it affects the communication overhead between the data storage units (e.g., the hard disks) and the computation units, e.g., CPUs and GPUs.
	All the simulations are conducted in Matlab.  The results in this section are obtained from 50 trials with different randomly generated tensors.

	\subsection{Results}
	Fig.~\ref{fig:10010mse} in Sec.~\ref{sec:intro} has shown the MSE performance of the algorithms in a relatively small-size example, where $I_n=I=100$, $F=10$ and the nonnegativity constraints are used in the algorithms.
	In that simulation, we use $|{\cal F}_n|=20$ so that every 500 iterations of the proposed algorithm compute a full MTTKRP.
	One can see that for this relatively easy case, all the algorithms can reach a good estimation accuracy for the latent factors.
	Nevertheless, the proposed methods exhibit remarkably higher efficiency.
	
\begin{figure}
	\centering
	\includegraphics[width=0.8\linewidth]{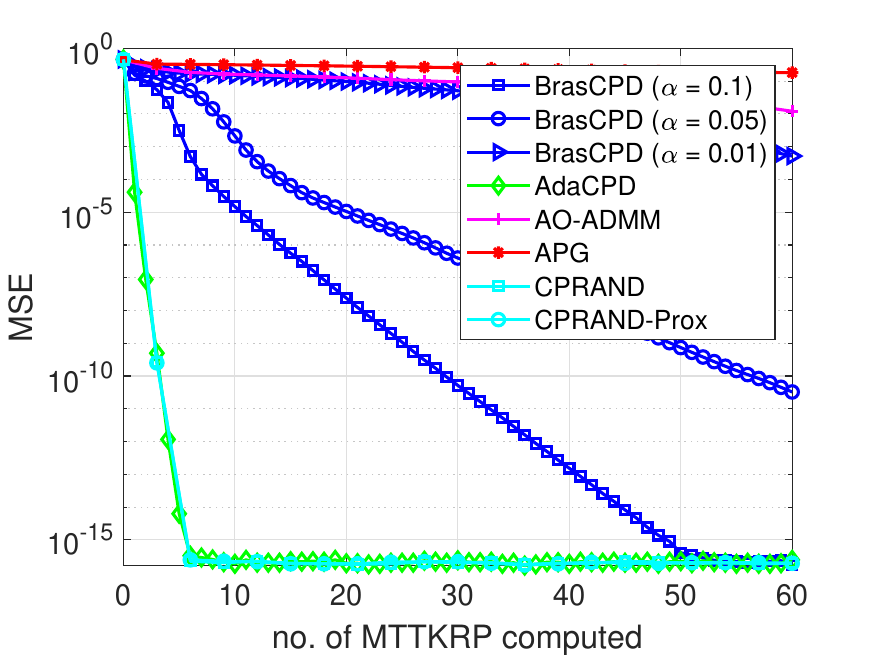}
	\caption{\black Median of MSEs against MTTKRP. $I_1=I_2=I_3=300$ and $F=10$. $\A_{(n)}\geq \bm 0$.}
	\label{fig:rank10_mttkrp}
	\end{figure}

	Fig.~\ref{fig:rank10_mttkrp} shows the MSEs of the estimated latent factors by the algorithms {\black under $I=300$ and $F=10$ benchmarked by more baselines.} The result is the \textit{median} of 50 Monte Carlo trials; we use median here since mean is dominated by outlying trials, even if there is only one outlying trial.
	We set $|{\cal F}_n|=18$ so that the proposed algorithms use 5,000 iterations to compute a full MTTKRP.
	All the algorithms use nonnegativity constraints except \texttt{CPRAND}. {\black One can see that under this setting, the most competitive algorithms are  \texttt{CPRAND},  \texttt{CPRAND-Prox}, and  \texttt{AdaCPD}. }

\begin{figure}
	\centering
	\includegraphics[width=0.8\linewidth]{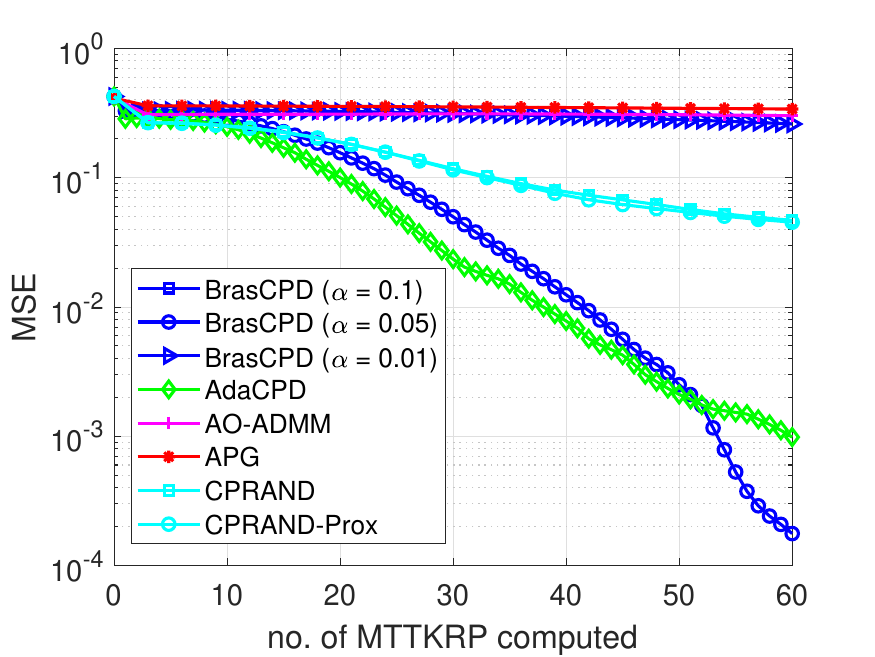}
	\caption{\black Median of MSEs against MTTKRP. $I_1=I_2=I_3=300$ and $F=200$. $\A_{(n)}\geq \bm 0$. }
	\label{fig:rank200_mttkrp}
\end{figure}

	{\black In Fig}.~\ref{fig:rank200_mttkrp}, {\black we increase the rank to be $F=200$.}
	There are several observations in order:
	First, the stochastic algorithms (i.e., \texttt{BrasCPD}, \texttt{AdaCPD}, {\black \texttt{CPRAND} and \texttt{CPRAND-Prox}}) are much more efficient relative to the deterministic algorithms (\texttt{AO-ADMM} and \texttt{APG}). After {\black 60} MTTKRPs computed, the stochastic algorithms often have reached a reasonable level of MSE.
	This is indeed remarkable, since {\black 60} MTTKRPs are roughly equivalent to {\black 20} iterations of \texttt{AO-ADMM} and \texttt{APG}.
	Second, {\black \texttt{CPRAND} and \texttt{CPRAND-Prox} are not as competitive in this high-rank regime}. In particular, \texttt{BrasCPD} with $\alpha=0.1$ gives the most promising performance.
	However, the performance of \texttt{BrasCPD} is affected a bit significantly by the parameters $\alpha$.
	One can see that using $\alpha=0.05$ and $\alpha=0.01$, the algorithm does not give promising results under this setting.
	Third, \texttt{AdaCPD} yields the second lowest MSEs {\black after 60 MTTKRPs}---while not using any parameter tuning. 

	\begin{figure}
	\centering
	\includegraphics[width=.85\linewidth]{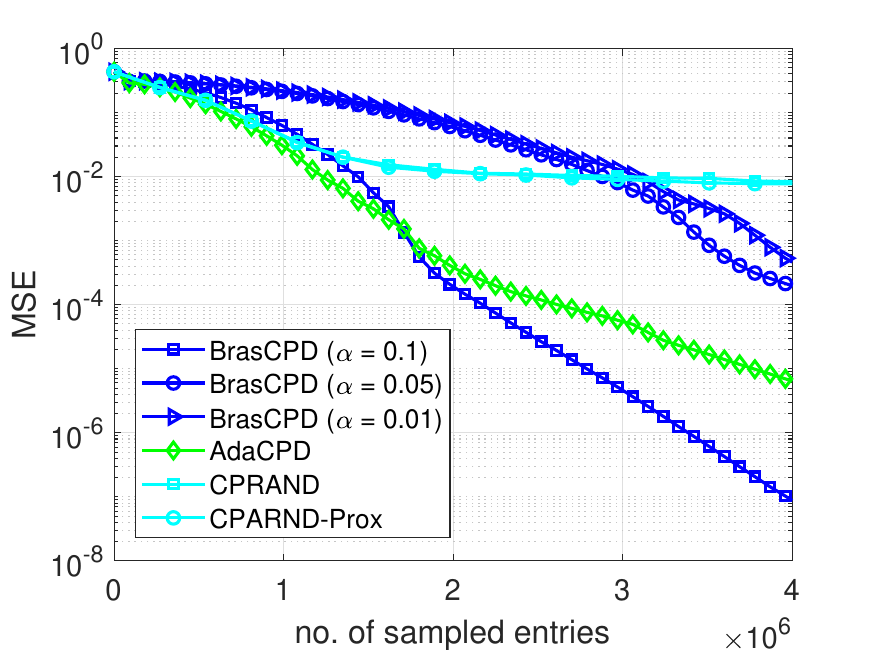}
	\caption{{\black MSE against the number of sampled entries for $I_1=I_2=I_3=300$, $F=100$.}}
	\label{fig:sampled}
\end{figure}

\begin{figure}
	\centering
	\includegraphics[width=.85\linewidth]{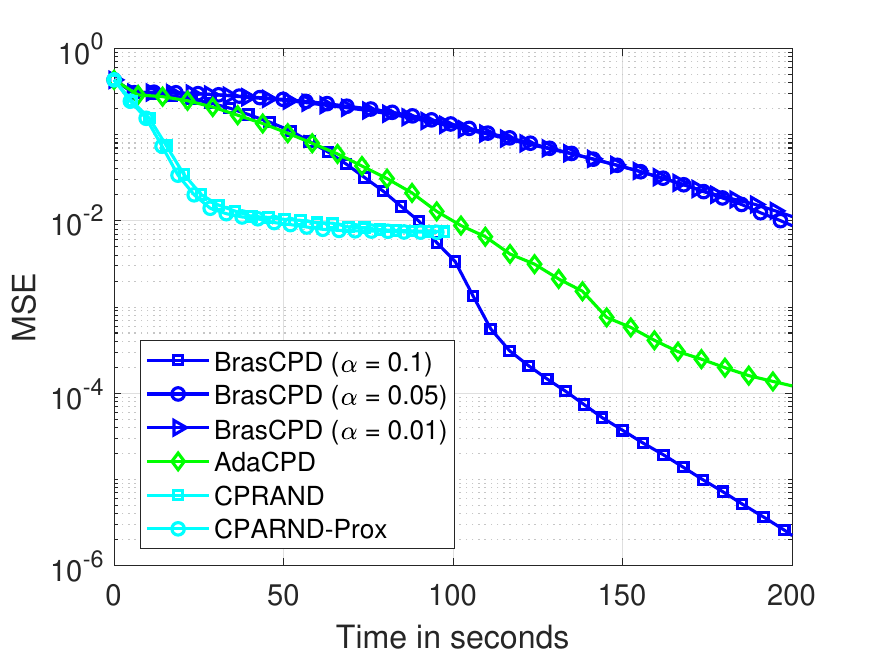}
	\caption{{\black MSE against runtime (sec.) for $I_1=I_2=I_3=300$, $F=100$.}}
	\label{fig:time_repeat}
\end{figure}
	
	Figs~\ref{fig:sampled}-\ref{fig:time_repeat} show the MSEs of the stochastic algorithms (i.e., \texttt{BrasCPD}, \texttt{AdaCPD}, {\black \texttt{CPRAND} and \texttt{CPRAND-Prox}}) against the number of sampled entries and runtime, respectively. {\black Here, we set $I_1=I_2=I_3=300$ and $F=100$.} One can see that  \texttt{BrasCPD} and  \texttt{AdaCPD} do not use many data samples to reach a good MSE level. This indicates that the communication overhead performance of the proposed algorithms is promising.
		In terms of runtime, \texttt{CPRAND} {\black and \texttt{CPRAND-Prox}} start with quick decrease of the MSE---{\black they reach} MSE$\approx 10^{-2}$ within 50 seconds, while the proposed approaches reach this level of accuracy after 100 seconds. Nevertheless, the {\black MSEs of \texttt{CPRAND} and \texttt{CPRADN-Prox} are} somehow stuck at this level, but the proposed algorithms can reach a much better accuracy for estimating the latent factors.
		We would like to remark that the runtime performance of stochastic algorithms are affected by the programming language used (i.e., \texttt{Matlab} in this case).
		Typically, interpreted languages (e.g., \texttt{Matlab} and \texttt{Python}) are not specialized for handling ``for'' loops, which is heavily used in stochastic algorithms (especially when $|{\cal F}_n|$ is small). Hence, real-system implementations for these algorithms could be much faster.

	Table~\ref{tab:Fchange} shows the mean and median of the MSEs and cost values output by the algorithms when the tensor rank varies under $I=300$.
	All the algorithms are stopped after {\black 60} full MTTKRPs are used.
	One can see that \texttt{BrasCPD} exhibits a quite competitive MSE performance if a proper $\alpha$ is chosen, under the employed stepsize schedule in \eqref{eq:schedule}. However, one can see that when $F$ changes, there is a risk that \texttt{BrasCPD} runs into numerical issues and yields unbounded solutions.
	This suggests that \texttt{BrasCPD} may need extra care for tuning its stepsize.
    In principle, when the problem setting changes, the ``best'' $\alpha$ of \texttt{BrasCPD} also changes. Our experience is that when $|{\cal F}_n|$ increases, using a properly scaled up $\alpha$ may help accelerate convergence.
	On the other hand, \texttt{AdaCPD} always outputs reasonably good results.
	More importantly,  \texttt{AdaCPD} runs without tuning the stepsize parameters---which shows the power of the adaptive stepsize scheduling strategy.
	{\black Another remark is that \texttt{CPRAND} and \texttt{CPRAND-Prox} work well when $F=10$ and $F=50$. In particular, \texttt{CPRAND-Prox} largely outperforms \texttt{CPRAND} when $F=50$, showing the benefit of explicitly considering constraints.}

\begin{figure}
		\includegraphics[width=.9\linewidth]{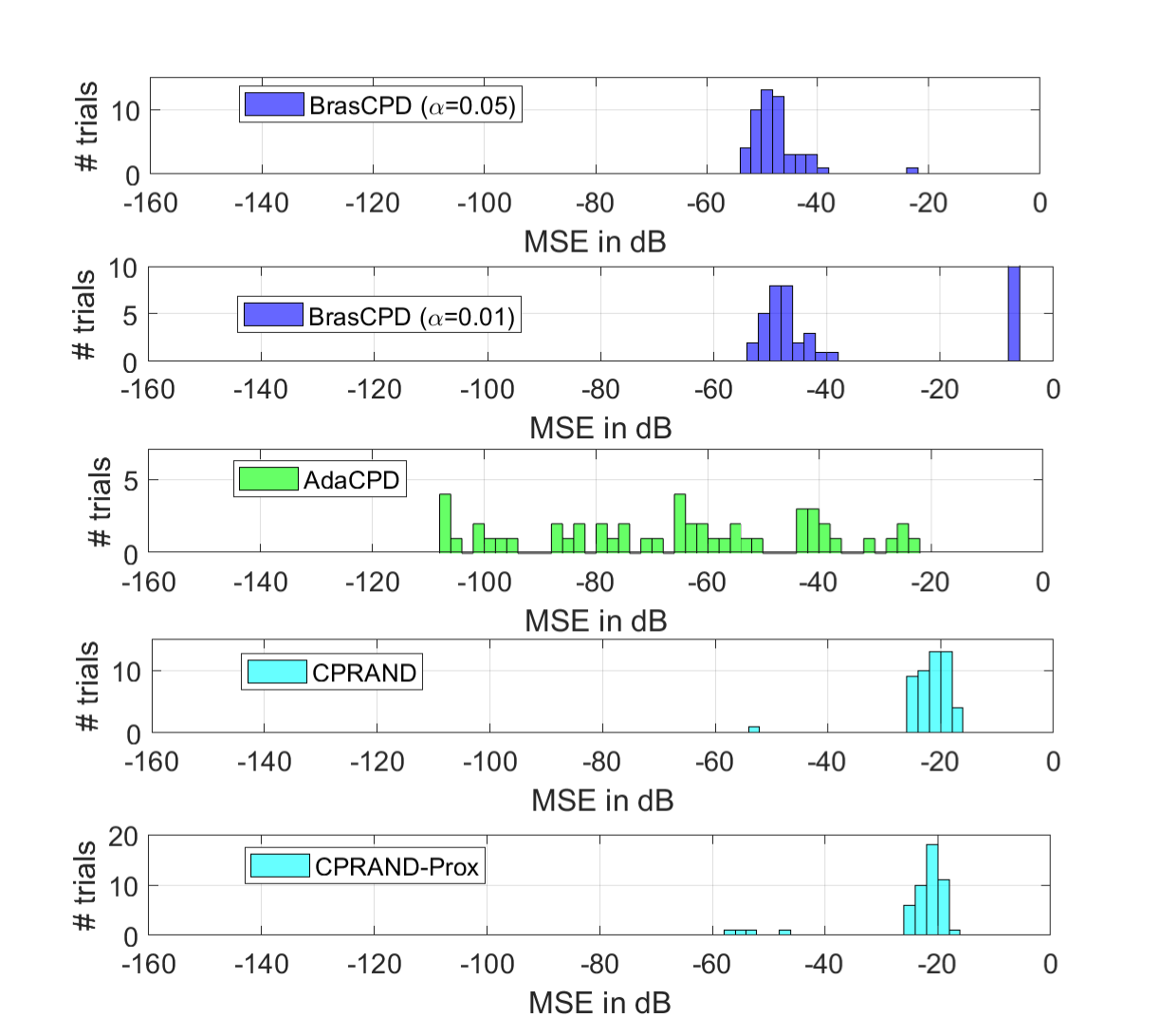}
		\caption{{\black Histograms of the algorithms; $I=300,F=100$.}}
		\label{fig:histogrami300f100edit5}
	\end{figure}

		Table~\ref{tab:Ichange} shows the performance under different $I$'s when $F=100$.  
		In general, when $I$ increases, the performance of all the algorithms improves---with a fixed $F$, a larger $I$ means more data and more ``degrees of freedom'' available, which normally leads to better performance.
		Again, \texttt{BrasCPD} with a proper $\alpha$ and \texttt{AdaCPD} in general outperform the baselines. One particular observation is that, although the mean and median MSEs of \texttt{AdaCPD} are both low, the median is sometimes much better than the mean (cf. the case when $I=400$), which indicates that there exist outlying trials (i.e., trials where \texttt{AdaCPD} does not produce very low MSE results). The median-mean gap is less often observed for other algorithms, e.g., \texttt{BrasCPD} and \texttt{CPRAND}. Fig.~\ref{fig:histogrami300f100edit5} may better illustrate the situation, where the histograms of the MSE (in dB) of the algorithms are shown. One can see that, although the {\black worst-case} result of \texttt{AdaCPD} is still acceptable (with MSE$<10^{-4}$), the MSE of \texttt{AdaCPD} clearly has a larger variance compared to \texttt{BrasCPD} with $\alpha=0.1$. This shows a trade-off between the easiness of stepsize scheduling and the risk of converging to less accurate solutions.
		
	{\black
		Table~\ref{tab:Ivaries_450} shows the median of the MSEs after 450 MTTKRPs of under same simulation setting (which is roughly equivalent to 150 iterations of AO-ADMM and APG). One can see that the performance of all the algorithms have improved. Note that due to resource limitations, we could not run 450 MTTKRPs for every table in this section and had to stop at 60 MTTKRPs. Hence, the capabilities of all the algorithms may not have fully shown up in those tables. Nonetheless, the 60-MTTKRP results can serve as a reference under limited time and computational resources.}

	\begin{table}[htbp]
			\centering
			\caption{MSEs of the estimated latent factors by the algorithms under different $F$; $I=300$; all the algorithms are stopped after computing 60 MTTKRPs. ``NaN'' means the algorithm outputs unbounded solutions. $\A_{(n)}\geq\bm 0$.}
			\resizebox{0.8\linewidth}{!}{
		\begin{tabular}{c|c|c|c|c|c|c}
			\hline
			\hline
			\multirow{2}[4]{*}{Algorithm} & \multicolumn{2}{c|}{\multirow{2}[4]{*}{Metric}} & \multicolumn{4}{c}{$F$} \\
			\cline{4-7}          & \multicolumn{2}{c|}{} & 10    & 50    & 100   & 200 \\
			\hline
			\hline
			\multirow{2}[4]{*}{BrasCPD $(\alpha=0.1)$} & \multirow{2}[4]{*}{MSE} & Mean  & 1.97E-16 & 1.98E-11 & 6.12E-10 & NaN \\
			\cline{3-7}          &       & Median & 1.70E-16 & 5.66E-12 & 3.82E-10 & NaN \\
			\hline
			\multirow{2}[4]{*}{BrasCPD $(\alpha=0.05)$} & \multirow{2}[4]{*}{MSE} & Mean  & 1.02E-10 & 1.33E-06 & 1.06E-04 & 4.08E-04 \\
			\cline{3-7}          &       & Median & 3.26E-11 & 7.20E-07 & 1.46E-05 & 1.77E-04 \\
			\hline
			\multirow{2}[4]{*}{BrasCPD $(\alpha=0.01)$} & \multirow{2}[4]{*}{MSE} & Mean  & 2.80E-03 & 0.1108 & 0.1968 & 0.2599 \\
			\cline{3-7}          &       & Median & 5.18E-04 & 0.1104 & 0.0000 & 0.2604 \\
			\hline
			\multirow{2}[4]{*}{AdaCPD } & \multirow{2}[4]{*}{MSE} & Mean  & 2.30E-16 & 1.27E-04 & 3.54E-04 & 0.0175 \\
			\cline{3-7}          &       & Median & 2.44E-16 & 5.43E-15 & 2.96E-07 & 9.86E-04 \\
			\hline
			\multirow{2}[4]{*}{AO-ADMM} & \multirow{2}[4]{*}{MSE} & Mean  & 2.25E-02 & 0.1900 & 0.2667 & 0.3018 \\
			\cline{3-7}          &       & Median & 1.21E-02 & 0.1886 & 0.2664 & 0.3018 \\
			\hline
			\multirow{2}[4]{*}{APG} & \multirow{2}[4]{*}{MSE} & Mean  & 1.80E-01 & 0.2993 & 0.3254 & 0.3399 \\
			\cline{3-7}          &       & Median & 1.83E-01 & 0.2995 & 0.3255 & 0.3398 \\
			\hline
			\multirow{2}[3]{*}{CPRAND} & \multirow{2}[3]{*}{MSE} & Mean  & 2.00E-16 & 0.0037 & 0.0082 & 0.0475 \\
			\cline{3-7}          &       & Median & 1.92E-16 & 0.0027 & 0.0075 & 0.0468 \\
			\hline
			\multirow{2}[2]{*}{CPRAND-Prox} & \multirow{2}[2]{*}{MSE} & Mean  & 2.02E-16 & 0.0026 & 0.0073 & 0.0459 \\
			\cline{3-7}          &       & Median & 1.92E-16 & 3.10E-11 & 0.0074 & 0.0453 \\
			\hline
			\hline
			\multirow{2}[3]{*}{BrasCPD $(\alpha=0.1)$} & \multirow{2}[3]{*}{Cost} & Mean  & 1.66E-19 & 2.63E-12 & 1.64E-10 & NaN \\
			\cline{3-7}          &       & Median & 4.31E-20 & 7.23E-13 & 1.03E-10 & NaN \\
			\hline
			\multirow{2}[4]{*}{BrasCPD $(\alpha=0.05)$} & \multirow{2}[4]{*}{Cost} & Mean  & 3.52E-12 & 1.84E-07 & 2.78E-05 & 3.01E-04 \\
			\cline{3-7}          &       & Median & 1.15E-12 & 1.00E-07 & 3.65E-06 & 8.99E-05 \\
			\hline
			\multirow{2}[4]{*}{BrasCPD $(\alpha=0.01)$} & \multirow{2}[4]{*}{Cost} & Mean  & 1.49E-04 & 0.0226 & 0.0698 & 0.1620 \\
			\cline{3-7}          &       & Median & 2.89E-05 & 0.0226 & 0.0701 & 0.1620 \\
			\hline
			\multirow{2}[4]{*}{AdaCPD } & \multirow{2}[4]{*}{Cost} & Mean  & 9.45E-31 & 4.95E-05 & 1.64E-04 & 0.0142 \\
			\cline{3-7}          &       & Median & 1.95E-31 & 8.86E-16 & 1.05E-07 & 8.43E-04 \\
			\hline
			\multirow{2}[4]{*}{AO-ADMM} & \multirow{2}[4]{*}{Cost} & Mean  & 1.10E-03 & 0.0361 & 0.0869 & 0.1791 \\
			\cline{3-7}          &       & Median & 8.10E-04 & 0.0359 & 0.0868 & 0.1790 \\
			\hline
			\multirow{2}[4]{*}{APG} & \multirow{2}[4]{*}{Cost} & Mean  & 2.77E-02 & 0.2665 & 0.5636 & 1.1709 \\
			\cline{3-7}          &       & Median & 2.79E-02 & 0.2659 & 0.5637 & 1.1715 \\
			\hline
			\multirow{2}[4]{*}{CPRAND} & \multirow{2}[4]{*}{Cost} & Mean  & 1.88E-31 & 1.79E-10 & 0.0023 & 0.0329 \\
			\cline{3-7}          &       & Median & 1.87E-31 & 1.59E-12 & 0.0027 & 0.0319 \\
			\hline
			\multirow{2}[4]{*}{CPRAND-Prox} & \multirow{2}[4]{*}{Cost} & Mean  & 1.89E-31 & 0.0168 & 0.2128 & 129.4668 \\
			\cline{3-7}          &       & Median & 1.88E-31 & 4.73E-12 & 0.1279 & 120.7575 \\
			\hline
			\hline
		\end{tabular}	\label{tab:Fchange}%
	}

	\end{table}
	
	\begin{table}[htbp]
			\centering
			\caption{ Performance of the algorithms under various $I$'s, $F=100$. $\A_{(n)}\geq\bm 0$; all the algorithms are stopped after computing 60 MTTKRPs. $\A_{(n)}\geq\bm 0$.  }
			\resizebox{0.8\linewidth}{!}{ 
				\begin{tabular}{c|c|c|c|c|c}
					\hline
					\hline
					\multirow{2}[4]{*}{Algorithm} & \multicolumn{2}{c|}{\multirow{2}[4]{*}{Metric}} & \multicolumn{3}{c}{$I$} \\
					\cline{4-6}          & \multicolumn{2}{c|}{} & 100   & 200   & 300 \\
					\hline
					\hline
					\multirow{2}[4]{*}{BrasCPD $(\alpha=0.1)$} & \multirow{2}[4]{*}{MSE} & Mean  & 0.1567 & 4.41E-04 & 6.12E-10 \\
					\cline{3-6}          &       & Median & 0.1586 & 8.54E-05 & 3.82E-10 \\
					\hline
					\multirow{2}[4]{*}{BrasCPD $(\alpha=0.05)$} & \multirow{2}[4]{*}{MSE} & Mean  & 0.2378 & 0.0297 & 1.06E-04 \\
					\cline{3-6}          &       & Median & 0.2364 & 0.0282 & 1.46E-05 \\
					\hline
					\multirow{2}[4]{*}{BrasCPD $(\alpha=0.01)$} & \multirow{2}[4]{*}{MSE} & Mean  & 0.2861 & 0.2786 & 0.1968 \\
					\cline{3-6}          &       & Median & 0.2857 & 0.2786 & 0.0000 \\
					\hline
					\multirow{2}[4]{*}{AdaCPD } & \multirow{2}[4]{*}{MSE} & Mean  & 0.1312 & 0.0040 & 3.54E-04 \\
					\cline{3-6}          &       & Median & 0.1281 & 1.66E-04 & 2.96E-07 \\
					\hline
					\multirow{2}[4]{*}{AO-ADMM} & \multirow{2}[4]{*}{MSE} & Mean  & 0.2396 & 0.2586 & 0.2667 \\
					\cline{3-6}          &       & Median & 0.2395 & 0.2585 & 0.2664 \\
					\hline
					\multirow{2}[4]{*}{APG} & \multirow{2}[4]{*}{MSE} & Mean  & 0.2961 & 0.3162 & 0.3254 \\
					\cline{3-6}          &       & Median & 0.2961 & 0.3163 & 0.3255 \\
					\hline
					\multirow{2}[3]{*}{CPRAND} & \multirow{2}[3]{*}{MSE} & Mean  & 0.1901 & 0.0139 & 0.0082 \\
					\cline{3-6}          &       & Median & 0.1914 & 0.0134 & 0.0075 \\
					\hline
					\multirow{2}[2]{*}{CPRAND-Prox} & \multirow{2}[2]{*}{MSE} & Mean  & 0.1873 & 0.0141 & 0.0073 \\
					\cline{3-6}          &       & Median & 0.1879 & 0.0146 & 0.0074 \\
					\hline
					\hline
					\multirow{2}[3]{*}{BrasCPD $(\alpha=0.1)$} & \multirow{2}[3]{*}{Cost} & Mean  & 0.0561 & 1.39E-04 & 1.64E-10 \\
					\cline{3-6}          &       & Median & 0.0556 & 2.22E-05 & 1.03E-10 \\
					\hline
					\multirow{2}[4]{*}{BrasCPD $(\alpha=0.05)$} & \multirow{2}[4]{*}{Cost} & Mean  & 0.0752 & 0.0119 & 2.78E-05 \\
					\cline{3-6}          &       & Median & 0.0744 & 0.0115 & 3.65E-06 \\
					\hline
					\multirow{2}[4]{*}{BrasCPD $(\alpha=0.01)$} & \multirow{2}[4]{*}{Cost} & Mean  & 0.1019 & 0.0881 & 0.0698 \\
					\cline{3-6}          &       & Median & 0.1010 & 0.0881 & 0.0701 \\
					\hline
					\multirow{2}[4]{*}{AdaCPD } & \multirow{2}[4]{*}{Cost} & Mean  & 0.0499 & 0.0018 & 1.64E-04 \\
					\cline{3-6}          &       & Median & 0.0499 & 5.01E-05 & 1.05E-07 \\
					\hline
					\multirow{2}[4]{*}{AO-ADMM} & \multirow{2}[4]{*}{Cost} & Mean  & 0.0736 & 0.0833 & 0.0869 \\
					\cline{3-6}          &       & Median & 0.0735 & 0.0833 & 0.0868 \\
					\hline
					\multirow{2}[4]{*}{APG} & \multirow{2}[4]{*}{Cost} & Mean  & 0.4248 & 0.5249 & 0.5636 \\
					\cline{3-6}          &       & Median & 0.4251 & 0.5251 & 0.5637 \\
					\hline
					\multirow{2}[4]{*}{CPRAND} & \multirow{2}[4]{*}{Cost} & Mean  & 0.0585 & 0.0046 & 0.0023 \\
					\cline{3-6}          &       & Median & 0.0575 & 0.0048 & 0.0027 \\
					\hline
					\multirow{2}[4]{*}{CPRAND-Prox} & \multirow{2}[4]{*}{Cost} & Mean  & 15.0076 & 1.2778 & 0.2128 \\
					\cline{3-6}          &       & Median & 12.3764 & 0.9751 & 0.1279 \\
					\hline
					\hline
				\end{tabular}	\label{tab:Ichange}
		}
	\end{table}%

	\begin{table}[htbp]
	\centering
	\caption{Median of the MSEs under various $I$'s, $F=100$. $\bm A_n \ge 0$; all the algorithms are stopped after computing 450 MTTKRPs.}
				\resizebox{0.8\linewidth}{!}{ 
	\begin{tabular}{c|c|c|c|c}
	\hline
	\hline
	\multirow{2}[4]{*}{Algorithm} & \multirow{2}[4]{*}{Metric} & \multicolumn{3}{c}{$I$} \\
	\cline{3-5}          &       & 100   & 200   & 300 \\
	\hline
	\hline
	BrasCPD $(\alpha=0.1)$ & MSE   & 3.76E-06 & 4.74E-17 & 9.56E-17 \\
	\hline
	BrasCPD $(\alpha=0.05)$ & MSE   & 0.0016 & 1.01E-14 & 8.75E-17 \\
	\hline
	BrasCPD $(\alpha=0.01)$ & MSE   & 0.1957 & 6.13E-04 & 8.44E-08 \\
	\hline
	AdaCPD  & MSE   & 9.51E-05 & 1.60E-07 & 2.43E-06 \\
	\hline
	AO-ADMM & MSE   & 3.45E-04 & 1.59E-04 & 1.44E-04 \\
	\hline
	APG   & MSE   & 0.2615 & 0.2873 & 0.3006 \\
	\hline
	CPRAND & MSE   & 0.0230 & 0.0070 & 0.0073 \\
	\hline
	CPRAND-Prox & MSE   & 0.0260 & 0.0069 & 0.0065 \\
	\hline
	\hline
\end{tabular}  	\label{tab:Ivaries_450}}
\end{table}%

	\begin{table}[htbp]
			\centering
			\caption{Performance of the algorithms under various SNRs;  all the algorithms are stopped after computing 60 MTTKRPs. $I_1=I_2=I_3=100$, $F=20$. $\A_{(n)}\geq\bm 0$.}
			\resizebox{0.85\linewidth}{!}{
			\begin{tabular}{c|c|c|c|c|c|c}
				\hline
				\hline
				\multirow{2}[4]{*}{Algorithm} & \multicolumn{2}{c|}{\multirow{2}[4]{*}{Metric}} & \multicolumn{4}{c}{SNR} \\
				\cline{4-7}          & \multicolumn{2}{c|}{} & 10    & 20    & 30    & 40 \\
				\hline
				\hline
				\multirow{2}[4]{*}{BrasCPD $(\alpha=0.1)$} & \multirow{2}[4]{*}{MSE} & Mean  & 0.0433 & 0.0240 & 0.0271 & 0.0281 \\
				\cline{3-7}          &       & Median & 0.0453 & 0.0195 & 0.0263 & 0.0257 \\
				\hline
				\multirow{2}[4]{*}{BrasCPD $(\alpha=0.05)$} & \multirow{2}[4]{*}{MSE} & Mean  & 0.1168 & 0.1135 & 0.1133 & 0.1093 \\
				\cline{3-7}          &       & Median & 0.1232 & 0.1149 & 0.1121 & 0.1090 \\
				\hline
				\multirow{2}[4]{*}{BrasCPD $(\alpha=0.01)$} & \multirow{2}[4]{*}{MSE} & Mean  & 0.2135 & 0.2114 & 0.2104 & 0.2087 \\
				\cline{3-7}          &       & Median & 0.2141 & 0.2121 & 0.2102 & 0.2078 \\
				\hline
				\multirow{2}[4]{*}{AdaCPD } & \multirow{2}[4]{*}{MSE} & Mean  & 0.0179 & 0.0040 & 0.0017 & 1.39E-04 \\
				\cline{3-7}          &       & Median & 0.0168 & 0.0036 & 7.02E-04 & 1.24E-04 \\
				\hline
				\multirow{2}[4]{*}{AO-ADMM} & \multirow{2}[4]{*}{MSE} & Mean  & 0.1012 & 0.0925 & 0.0874 & 0.0905 \\
				\cline{3-7}          &       & Median & 0.1009 & 0.0921 & 0.0870 & 0.0899 \\
				\hline
				\multirow{2}[4]{*}{APG} & \multirow{2}[4]{*}{MSE} & Mean  & 0.2326 & 0.2291 & 0.2282 & 0.2284 \\
				\cline{3-7}          &       & Median & 0.2319 & 0.2286 & 0.2279 & 0.2291 \\
				\hline
				\multirow{2}[3]{*}{CPRAND} & \multirow{2}[3]{*}{MSE} & Mean  & 0.2049 & 0.0088 & 0.0021 & 0.0020 \\
				\cline{3-7}          &       & Median & 0.2046 & 0.0077 & 7.36E-04 & 7.33E-05 \\
				\hline
				\multirow{2}[2]{*}{CPRAND-Prox} & \multirow{2}[2]{*}{MSE} & Mean  & 0.1737 & 0.0108 & 0.0024 & 0.0017 \\
				\cline{3-7}          &       & Median & 0.1715 & 0.0075 & 7.32E-04 & 7.44E-05 \\
				\hline
				\hline
			\end{tabular} 	\label{tab:snr_nn}}
	\end{table}%

	\begin{table}[htbp]
			\centering
			\caption{Performance of the algorithms under various SNRs after computing 30 MTTKRPs. $I_1=I_2=I_3=100$, $F=20$. ${\bm 1}^\T\A_{(n)}=\rho\bm 1^\T$, $\A_{(n)}\geq\bm 0$. $\rho=100$. }
			\resizebox{0.85\linewidth}{!}{    
				\begin{tabular}{c|c|c|c|c|c|c}
					\hline
					\hline
					\multirow{2}[4]{*}{Algorithm} & \multicolumn{2}{c|}{\multirow{2}[4]{*}{Metric}} & \multicolumn{4}{c}{SNR} \\
					\cline{4-7}          & \multicolumn{2}{c|}{} & 10    & 20    & 30    & 40 \\
					\hline
					\hline
					\multirow{2}[4]{*}{BrasCPD $(\alpha=0.1)$} & \multirow{2}[4]{*}{MSE} & Mean  & 0.3946 & 0.3950 & 0.3939 & 0.4032 \\
					\cline{3-7}          &       & Median & 0.3987 & 0.3993 & 0.3942 & 0.4062 \\
					\hline
					\multirow{2}[4]{*}{BrasCPD $(\alpha=0.05)$} & \multirow{2}[4]{*}{MSE} & Mean  & 0.0705 & 0.0067 & 6.80E-04 & 6.82E-05 \\
					\cline{3-7}          &       & Median & 0.0705 & 0.0067 & 6.81E-04 & 6.80E-05 \\
					\hline
					\multirow{2}[4]{*}{BrasCPD $(\alpha=0.01)$} & \multirow{2}[4]{*}{MSE} & Mean  & 0.0169 & 0.0054 & 0.0017 & 0.0026 \\
					\cline{3-7}          &       & Median & 0.0156 & 0.0017 & 2.34E-04 & 7.75E-05 \\
					\hline
					\multirow{2}[4]{*}{AdaCPD } & \multirow{2}[4]{*}{MSE} & Mean  & 0.0202 & 0.0019 & 3.94E-04 & 6.13E-05 \\
					\cline{3-7}          &       & Median & 0.0193 & 0.0019 & 3.95E-04 & 6.11E-05 \\
					\hline
					\multirow{2}[4]{*}{AO-ADMM} & \multirow{2}[4]{*}{MSE} & Mean  & 0.0884 & 0.0759 & 0.0824 & 0.0772 \\
					\cline{3-7}          &       & Median & 0.0881 & 0.0763 & 0.0823 & 0.0805 \\
					\hline
					\multirow{2}[4]{*}{APG} & \multirow{2}[4]{*}{MSE} & Mean  & 0.2186 & 0.2149 & 0.2144 & 0.2146 \\
					\cline{3-7}          &       & Median & 0.2188 & 0.2152 & 0.2157 & 0.2148 \\
					\hline
					\multirow{2}[3]{*}{CPRAND} & \multirow{2}[3]{*}{MSE} & Mean  & 0.1937 & 0.0097 & 0.0034 & 0.0032 \\
					\cline{3-7}          &       & Median & 0.1944 & 0.0077 & 7.44E-04 & 7.37E-05 \\
					\hline
					\multirow{2}[2]{*}{CPRAND-Prox} & \multirow{2}[2]{*}{MSE} & Mean  & 0.1509 & 0.0670 & 0.0599 & 0.0602 \\
					\cline{3-7}          &       & Median & 0.1519 & 0.0663 & 0.0591 & 0.0582 \\
					\hline
					\hline
				\end{tabular}
				\label{tab:simplex}}%
		\vspace{-.5cm}
	\end{table}%

	Tables~\ref{tab:snr_nn}-\ref{tab:simplex} show the estimation accuracy of the latent factors by the algorithms under different {\sf SNR}s. {\black Here, we use a low-rank case where $I_1=I_2=I_3=100$ and $F=20$, under which \texttt{CPRAND} and \texttt{CPRAND-Prox} are more competitive}.
	In a noisy environment, the ability of handling constraints/regularizations is essential for a CPD algorithm, since prior information on the latent factors can help improve estimation accuracy. Table~\ref{tab:snr_nn} and Table~\ref{tab:simplex} test the cases where $\A_{(n)}$ is elementwise nonnegative and the columns of $\A_{(n)}$ reside in a scaled version of the probability simplex, respectively. One can see from the two tables that both \texttt{BrasCPD} (with a proper $\alpha$) and \texttt{AdaCPD} work very well. 
	In Table~\ref{tab:simplex}, one can see that \texttt{BrasCPD} again shows its sensitivity to the choice of $\alpha$, with $\alpha=0.1$ and $0.05$ actually not working. 
	{\black One can also see that, in the low-SNR regime (SNR=$10$ and $20$dB), \texttt{BrasCPD} with proper $\alpha$ and \texttt{AdaCPD} outperform the baselines.}
	{\black \texttt{CPRAND-Prox} also works well for the nonnegativity constraint case (especially for SNR=$30$ and 40dB), but not as promising for the simplex constraints.}

	\subsection{Real-Data Experiment}
	In this subsection, we test our algorithm on a constrained tensor decomposition problem; i.e., we apply the proposed \texttt{BrasCPD} and \texttt{AdaCPD} to factor hyperspectral images. Hyperspectral images (HSIs) are special images with pixels measured at a large number of wavelengths. Hence, an HSI is usually stored as a third-order tensor with two spatial coordinates and one spectral coordinate. HSIs are dense tensors and thus are suitable for testing the proposed algorithms.
	We use sub-images of the Indian Pines dataset that has a size of $145\times 145\times 220$ and the Pavia University dataset\footnote{Both datasets are available online: \url{http://www.ehu.eus/ccwintco/index.php/Hyperspectral_Remote_Sensing_Scenes}} that has a size of $610\times 340\times 103$. {\black We apply the stochastic algorithms to the datasets by fixing $|{\cal F}_n|=500$ in this section.}
	\begin{table}[htbp]
		\centering
		{\black
		\caption{Performance of the algorithms on the Indian Pines dataset under different $F$'s. }
		\resizebox{0.85\linewidth}{!}{\huge
			\begin{tabular}{c|c|c|c|c|c}
				\hline
				\hline
				\multirow{2}[4]{*}{Algorithm} & \multirow{2}[4]{*}{Metric} & \multicolumn{4}{c}{$F$} \\
				\cline{3-6}          &       & 10    & 20    & 30    & 40 \\
				\hline
				\hline
				BrasCPD ($\alpha$=4) & Cost  & 6.82E-04 & 4.68E-04 & 3.46E-04 & 3.58E-04 \\
				\hline
				BrasCPD ($\alpha$=3) & Cost  & 6.86E-04 & 4.52E-04 & 3.54E-04 & 3.35E-04 \\
				\hline
				BrasCPD ($\alpha$=2) & Cost  & 6.88E-04 & 6.11E-04 & 5.38E-04 & 4.11E-04 \\
				\hline
				AdaCPD  & Cost  & 6.23E-04 & 4.56E-04 & 5.40E-04 & 5.23E-04 \\
				\hline
				AO-ADMM & Cost  & 7.70E-04 & 5.28E-04 & 5.07E-04 & 4.91E-04 \\
				\hline
				APG   & Cost  & 0.0019 & 0.0019 & 0.0018 & 0.0019 \\
				\hline
				CPRAND & Cost  & 6.74E-04 & 4.82E-04 & 4.35E-04 & 4.08E-04 \\
				\hline
				CPRAND-Prox & Cost  & 0.1116 & 0.0021 & 0.0020 & 0.0021 \\
				\hline
				\hline
		\end{tabular}}%
		\label{tab:indianpines}%
	}
	\end{table}%
	
	\begin{table}[htbp]
		\centering
		{\black 
		\caption{Performance of the algorithms on the Pavia University dataset under different $F$'s. }
		\resizebox{0.85\linewidth}{!}{
			\begin{tabular}{c|c|c|c|c}
				\hline
				\hline
				\multirow{2}[4]{*}{Algorithm} & \multirow{2}[4]{*}{Metric} & \multicolumn{3}{c}{$F$} \\
				\cline{3-5}          &       & 50    & 100   & 200 \\
				\hline
				\hline
				BrasCPD ($\alpha$=4) & Cost  & 0.0031 & 0.0027 & 0.0013 \\
				\hline
				BrasCPD ($\alpha$=3) & Cost  & 0.0033 & 0.0053 & 0.0031 \\
				\hline
				BrasCPD ($\alpha$=2) & Cost  & 0.0044 & 0.0067 & 0.0059 \\
				\hline
				AdaCPD  & Cost  & 0.0022 & 0.0013 & 0.0008 \\
				\hline
				AO-ADMM & Cost  & 0.0378 & 0.0425 & 0.0053 \\
				\hline
				APG   & Cost  & 0.0074 & 0.0073 & 0.0031 \\
				\hline
				CPRAND & Cost  & 0.0033 & 0.0028 & 0.0034 \\
				\hline
				CPRAND-Prox & Cost  & 0.0103 & 0.0109 & 8.46E+03 \\
				\hline
				\hline
			\end{tabular}
			\label{tab:univer}
		}%
	   }%
	\end{table}%
		
	Tables~\ref{tab:indianpines}-\ref{tab:univer} show  the cost values of the nonnegativity constrained optimization algorithms under different ranks, after computing {\black 120} MTTKRPs for all three modes, which corresponds to {\black 120} iterations for \texttt{AO-ADMM} and \texttt{APG} (we use this ``all-mode MTTKRP'' in this section since the tensors are unsymmetrical and thus single-mode MTTKRPs cannot be directly translated to iterations in batch algorithms).
	One can see that the proposed algorithms show the same merits as we have seen in the simulations: \texttt{BrasCPD} can exhibit very competitive performance when $\alpha$ is properly chosen (e.g., when $F=10$ and $\alpha = 5$ for the Indian Pines dataset); in addition, \texttt{AdaCPD} gives consistently good performance without tuning the stepsize manually. 
	Particularly, on the Pavia University dataset, \texttt{AdaCPD} gives much lower cost values compared to other algorithms. {\black We also present the results of \texttt{CPRAND} and \texttt{CPRAND-Prox}. Note that since \texttt{CPRAND} does not have constraints, its cost value is naturally lower than the other methods. Hence, this baseline is only for reference. Both \texttt{CPRAND} and \texttt{CPRAND-Prox} work reasonably well for the two datasets, especially on Indian Pines under low rank. However, we note that for Pavia University, \texttt{CPRAND-Prox} does not converge well for the $F=200$ case.}
	
\begin{figure}[H]
	\centering
	\includegraphics[width=.8\linewidth]{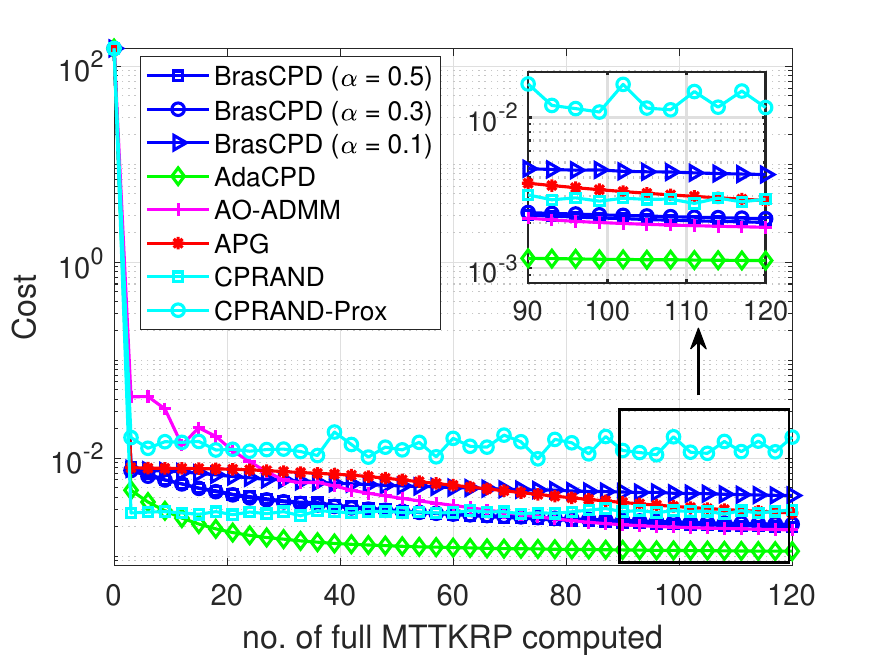}
	\caption{\black Number of all-mode MTTKRPs v.s. cost values output by the algorithms when applied to the Pavia University dataset. $F=100$. Nonnegativity constraint is added .}
	\label{fig:realdata}
\end{figure}

	Fig.~\ref{fig:realdata} shows how the cost values change along with the iterations on the Pavia University data using $F=100$. One can see that \texttt{BrasCPD} ($\alpha=0.5$) and \texttt{AdaCPD} reduce the cost value quickly in this case. After 120 iterations (equivalent to 120 all-mode full MTTKRPs), the batch algorithm \texttt{APG} eventually reaches the same cost value level of those of \texttt{BrasCPD} ($\alpha=0.5$) and \texttt{AdaCPD}.

	\section{Conclusion}
	To conclude, we proposed a block-randomized stochastic proximal gradient based CPD algorithmic framework for large-scale dense tensors.
	The framework works under a doubly stochastic manner, which randomly selects a mode and then samples a set of fibers for updating the associated latent factor.
	The framework has a series of nice features including being able to quickly improve estimation accuracy of the latent factors,
	being flexible with incorporating constraints and regularizations,
	and having rigorous convergence guarantees. 
	We also proposed a practical and effective adaptive stepsize scheduling method that is reminiscent of recent advances in neural network training algorithms.
	Simulations and real-data experiments show that the proposed algorithms outperform a number of state-of-art constrained CPD algorithms when dealing with large dense tensors.

	\appendices

	\section{Connection between $\nabla f(\bm \theta^{(r)})$ and $\bm G^{(r)}$}\label{app:G}
	Let $n$ be any integer from $\{1,...,N\}$, consider the following conditional expectation:
	\begin{align}
	& \overline{\bm G}_{(n)}^{(r)} = \mathbb{E}_{ \xi^{(r)}, \bm{\zeta}^{(r)} } \big[ {\bm G}_{(n)}^{(r)} \!~ | \!~ {\cal B}^{(r)} \big]
	\nonumber\\
	& \overset{(a)}{=} \mathbb{E}_{\xi^{(r)}} \left[  \frac{1}{{J_{\xi^{(r)}} \choose B}} \big( {\bm A}_{(\xi^{(r)})}^{(r)} {\bm H}_{(\xi^{(r)})}^\T {\bm H}_{(\xi^{(r)})} - {\bm X}_{(\xi^{(r)})}^\top {\bm H}_{(\xi^{(r)})} \big) \right] \nonumber\\
	& \overset{(b)}{=} \sum_{n'=1}^N \frac{\delta( n' - n )}{N {J_{n'} \choose B}} \big( {\bm A}_{(n')}^{(r)} {\bm H}_{(n')}^\T {\bm H}_{(n')} - {\bm X}_{(n')}^\top {\bm H}_{(n')} \big) \nonumber\\
	& = \frac{1}{N {J_{n} \choose B}} \big( {\bm A}_{(n)}^{(r)} {\bm H}_{(n)}^\T {\bm H}_{(n)} - {\bm X}_{(n)}^\top {\bm H}_{(n)} \big) \label{eq:fg}
	\end{align}
	where $\delta(\cdot)$ is the Dirac function. In the above chain, (a) is due to Fact~\ref{fact:unbias} and (b) is obtained by evaluating the expectation with respect to  the possible modes $\xi^{(r)} = n'$.  
	The last equality shows that $\overline{\bm G}^{(r)}_{(n)}$ is a scaled gradient of the objective function of \eqref{eq:lscri} 
	taken w.r.t.~${\bm A}_{(n)}^{(r)}$. The block sampling step together with fiber sampling entails us an easy way to estimate the \textit{full gradient w.r.t. all the latent factors} in an unbiased manner.

	\section{Proof of Proposition~\ref{prop:unconstrained}}\label{app:proof_unsconstrained}

	To show Proposition~\ref{prop:unconstrained}, we will need the following \cite[Lemma A.5]{mairal2010online}:
	\begin{Lemma}\label{lem:nnsequence}
		Let $\{ a_t \}_{t}$ and $\{ b_t \}_t$ be two nonnegative sequences such that $b_t$ is bounded, $\sum_{t=0}^\infty a_tb_t$ converges and $\sum_{t=0}^\infty a_t$ diverges, then we have
		\[   \displaystyle \liminf_{r \rightarrow \infty}  b_t =0.  \]
	\end{Lemma}

	Recall that $\xi^{(r)}$, $\bm{\zeta}^{(r)}$ are the random mode, fiber chosen at iteration $r$, respectively. 
	Under Assumption~\ref{as:boundedness}, we have $\| {\bm H}_{(\xi^{(r)})}^{(r)} \|_2^2 \leq L_{(\xi^{(r)})}^{(r)}$
	where ${\bm H}_{(\xi^{(r)})}^{(r)} = \odot_{n' =1,n' \neq \xi^{(r)}}^N \A_{(n')}^{(r)}$ and $L_{(\xi^{(r)})}^{(r)}<\infty$.
	{Combining with Fact~\ref{fact:Lcon}, we observe:
		\begin{align*}
		&f(\bm \theta^{(r+1)})-f(\bm \theta^{(r)}) \leq \left\langle  \nabla_{\A_{(\xi^{(r)})} } f(\bm \theta^{(r)}),\A^{(r+1)}_{(\xi^{(r)})} - \A^{(r)}_{(\xi^{(r)})} \right \rangle \\
		& \hspace{4.75cm} + \frac{L}{2}\left\|   \A^{(r+1)}_{(\xi^{(r)})} - \A^{(r)}_{(\xi^{(r)})} \right\|^2\\
		& =-\alpha^{(r)} \left\langle  \nabla_{\A_{(\xi^{(r)})} } f(\bm \theta^{(r)}), {\bm G}_{(\xi^{(r)})}^{(r)} \right \rangle + \frac{(\alpha^{(r)})^2 L}{2}\left\|   {\bm G}_{(\xi^{(r)})}^{(r)} \right\|^2,
		\end{align*}
		where we have denoted
		\[  L =\max_{r=0,\ldots,\infty} ~L_{(\xi^{(r)})}^{(r)}<\infty. \]
		
		Taking expectation conditioned on the filtration ${\cal B}^{(r)}$ and the chosen mode index $\xi^{(r)}$, we have
		\begin{align}
		&\mathbb{E}_{\bm\zeta^{(r)}} \left[f(\bm \theta^{(r+1)}) \!~ | \!~ {\cal B}^{(r)}, \xi^{(r)} \right]-f(\bm \theta^{(r)}) \nonumber\\
		& \leq -\alpha^{(r)} \left\| \nabla_{\A_{(\xi^{(r)})}} f(\bm \theta^{(r)})  \right\|^2 \nonumber\\
		&\quad\quad\quad +  \frac{(\alpha^{(r)})^2 L }{2} \mathbb{E}_{\bm\zeta^{(r)}} \left[ \left\| {\bm G}_{(\xi^{(r)})}^{(r)} \right\|^2 \!~ | \!~ {\cal B}^{(r)}, \xi^{(r)} \right] \nonumber\\
		&\leq -\alpha^{(r)} \left\| \nabla_{\A_{(\xi^{(r)})}} f(\bm \theta^{(r)})  \right\|^2+ \frac{(\alpha^{(r)})^2 L  M }{2}, \label{eq:suff_dec}
		\end{align}
		where
		the first inequality used the assumption
		that $L_{(\xi^{(r)})}^{(r)}\leq L$ and Fact~\ref{fact:unbias}, and the second inequality is  a consequence of Assumption~\ref{as:boundedness}, as we observe:
		\begin{equation}
		\begin{split}
		\left\| {\bm G}_{(\xi^{(r)})}^{(r)} \right\| & = \frac{1}{B} \Big\| {\bm A}_{(\xi^{(r)})}^{(r)} ({\bm H}_{(\xi^{(r)})}^{(r)} (\bm\zeta^{(r)}) )^\T {\bm H}_{(\xi^{(r)})}^{(r)}(\bm\zeta^{(r)}) \\
		& \hspace{2cm} - {\bm X}_{(\xi^{(r)})}^\T (\bm\zeta^{(r)}) {\bm H}_{(\xi^{(r)})}^{(r)} (\bm\zeta^{(r)}) \Big\|.
		\end{split}
		\end{equation}
		As ${\bm X}_{(n)}$ is bounded for all $n$, and all the ${\bm A}_{(n)}^{(r)}$ are bounded under Assumption~\ref{as:boundedness},
		we have $\| {\bm G}_{(\xi^{(r)})}^{(r)} \|^2 \leq M$ for all $n,r$ and some $M < \infty$.
		Taking the expectation w.r.t.~$\xi^{(r)}$ yields
		\begin{align}
		&{\black \mathbb{E}_{\xi^{(r)},\zeta^{(r)}}} \left[f(\bm \theta^{(r+1)}) \!~ | \!~ {\cal B}^{(r)} \right]-f(\bm \theta^{(r)}) \nonumber\\
		&\leq -\alpha^{(r)} \mathbb{E}_{\xi^{(r)}}\left[ \left\| \nabla_{\A_{(\xi^{(r)})}} f(\bm \theta^{(r)})  \right\|^2 \right]+ \frac{(\alpha^{(r)})^2ML}{2}. \label{eq:suff_dec_v2}
		\end{align}
		Note that $\mathbb{E}_{\xi^{(r)}}[ \| \nabla_{\A_{(\xi^{(r)})}}f(\bm \theta^{(r)})\|^2 ] = \| \nabla f(\bm \theta^{(r)}) \|^2$.
		Taking the total expectation (w.r.t.~all random variables in ${\cal B}^{(r)}$) gives
		\begin{align}
		&\mathbb{E}\left[f(\bm \theta^{(r+1)})\right]-\mathbb{E}\left[f(\bm \theta^{(r)})\right] \nonumber\\
		&\leq -\alpha^{(r)} \mathbb{E}\left[  \left\| \nabla  f(\bm \theta^{(r)})  \right\|^2  \right]+ \frac{(\alpha^{(r)})^2ML}{2}. \label{eq:suff_dec_v3}
		\end{align}
		Summing up \eqref{eq:suff_dec_v3} from $t=0$ to $t=r$, we have
		\begin{align}
		&\mathbb{E}\left[f(\bm \theta^{(t+1)})\right]-f(\bm \theta^{(0)}) \nonumber\\
		&\leq \sum_{t=0}^r-\alpha^{(t)} \mathbb{E}\left[ \left\| \nabla  f(\bm \theta^{(t)})  \right\|^2 \right] + \sum_{t=0}^r\frac{(\alpha^{(t)})^2ML}{2}. \nonumber
		\end{align}
		Taking $r\rightarrow \infty$, the above implies that
		\begin{align}
		& \sum_{r=0}^\infty \alpha^{(r)} \mathbb{E}\left[ \left\| \nabla  f(\bm \theta^{(r)})  \right\|^2 \right]\nonumber\\
		&\quad\quad\leq f(\bm \theta^{(0)})-f(\bm \theta^{(\star)}) + \sum_{r=0}^\infty\frac{(\alpha^{(r)})^2ML}{2},
		\end{align}
		where we have used $f( \bm{\theta} ) \geq f(\bm \theta^{(\star)})$, and $f(\bm \theta^{(\star)})$ denotes the global optimal value.
		Note that the right hand side above is bounded from above because
		$  \sum_{r=0}^\infty (\alpha^{(r)})^2<\infty$.
		Hence, using Lemma~\ref{lem:nnsequence}
		we conclude:
		\[ \displaystyle \liminf_{r \rightarrow \infty}  \mathbb{E} \big[ \big\| \nabla f(\bm \theta^{(r)})  \big\|^2  \big] = 0. \]}

	\section{Proof of Proposition~\ref{prop:prox}}
	\subsection{Preliminaries}
	For the constrained case, let us denote $\Phi(\bm \theta)=f(\bm \theta)+\sum_{n=1}^N h_n(\bm \theta)$ as the objective function.
	Unlike the unconstrained case where we measure convergence via observing if the gradient vanishes, the optimality condition of the constrained case is a bit more complicated.
	Consider the following optimization problem
	\[ \minimize_{\bm \theta}~f(\bm \theta) + h(\bm \theta), \]
	where $f(\bm \theta)$ is continuously differentiable while $h$ is convex but possibly nonsmooth.
	The deterministic proximal gradient algorithm for handling this problem is as follows:
	\[     \bm \theta^{(r+1)}\leftarrow {\sf Prox}_h\left(\bm \theta^{(r)} - \alpha^{(r)} \nabla f(\bm \theta^{(r)})\right).     \]
	Define $\bm P^{(r)} = \frac{1}{\alpha^{(r)}}\left( \bm \theta^{(r+1)} - \bm \theta^{(r)}  \right)$, the update can also be represented as
	$ \bm \theta^{(r+1)}\leftarrow \bm \theta^{(r)}-\alpha^{(r)}\bm P^{(r)}, $
	which is analogous to the gradient descent algorithm. It can be shown that $\bm P^{(r)}=\bm 0$ implies that the necessary optimality condition is satisfied, and thus $\bm P^{(r)}$ can be considered as a ``generalized gradient''.	
	In the multi-block setting of \eqref{eq:lscri_reg}, we define:
	\begin{align*}
	&\bm P_{(n)}^{(r)}\\
	&= \frac{1}{\alpha^{(r)}} \left(  \A_{(n)}^{(r+1)} - {\sf Prox}_{h_n}\left(  \A_{(n)}^{(r)} - \alpha^{(r)}\nabla_{\A_{(n)}} f(\bm \theta^{(r)})  \right) \right). 
	\end{align*}   
	To show that the \texttt{BrasCPD} algorithm finds a stationary point, our goal is to show the subsequence convergence of $\mathbb{E}\left[\bm P_{(n)}^{(r)} \right]$ to zero for all $n$ as $r \rightarrow \infty$.

	\subsection{Proof}
	Our update is equivalent to the following:
	\begin{align}
	\A_{(n)}^{(r+1)}\leftarrow \arg\min_{\A_{(n)}}~&\left\langle \G_{(n)}^{(r)}, \A_{(n)} - \A_{(n)}^{(r)} \right\rangle \label{eq:prob_prox}\\
	& + \frac{1}{2\alpha^{(r)}}\left\|\A_{(n)} -\A_{(n)}^{(r)}\right\|^2 + h_n(\A_{(n)})  \nonumber
	\end{align}
	for a randomly selected $n$, i.e., the above is the proximal operator.
	For a given $\xi^{(r)}$, we have
	\begin{align*}
	&h_{\xi^{(r)}}\left(  \A_{(\xi^{(r)})}^{(r+1)} \right) - h_{\xi^{(r)}}\left(  \A_{(\xi^{(r)})}^{(r)} \right) \\
	\leq&  - \left\langle \G_{(\xi^{(r)})}^{(r)},  \A_{(\xi^{(r)})}^{(r+1)} - \A_{(\xi^{(r)})}^{(r)} \right\rangle - \frac{1}{2\alpha^{(r)}}\left\|\A_{(\xi^{(r)})}^{(r+1)} - \A_{(\xi^{(r)})}^{(r)} \right\|^2
	\end{align*}
	by the optimality of $ \A_{(\xi^{(r)})}^{(r+1)} $ for solving Problem~\eqref{eq:prob_prox}.
	
	By the block  Lipschitz continuity of the smooth part (cf. Fact~\ref{fact:Lcon}), we have
	\begin{align*}
	f(\bm \theta^{(r+1)}) - f(\bm \theta^{(r)}) &\leq \left \langle  \nabla_{\A_{(\xi^{(r)})}}~f(\bm \theta^{(r)}),   \A_{(\xi^{(r)})}^{(r+1)} - \A_{(\xi^{(r)})}^{(r)}   \right \rangle\\
	& \quad  + \frac{L_{(\xi^{(r)})}^{(r)}}{2}\left\|\A_{(\xi^{(r)})}^{(r+1)} - \A_{(\xi^{(r)})}^{(r)} \right\|^2,
	\end{align*}
	where $f$ denotes the smooth part in the objective function and 
	\[   L_{(\xi^{(r)})}^{(r)} =\lambda_{\max}\left( \left(\H_{(\xi^{(r)})}^{(r)}\right)^\T\H_{(\xi^{(r)})}^{(r)} \right) \leq L.  \]
	Combining the two inequalities, we have
	\begin{align}\label{eq:descent}
	\Phi(\bm \theta^{(r+1)})&\leq \Phi(\bm \theta^{(r)})\nonumber\\ &\quad\quad-{\alpha^{(r)} \left\langle \nabla_{\A_{(\xi^{(r)})}} f(\bm \theta^{(r)})-\G_{(\xi^{(r)})}^{(r)},\bm p_{(\xi^{(r)})}^{(r)}\right\rangle}\nonumber\\
	&\quad\quad+ \left(\frac{L(\alpha^{(r)})^2}{2} - \frac{\alpha^{(r)}}{2}\right)\|\bm p_{(\xi^{(r)})}^{(r)}\|^2,
	\end{align}
	where we have defined:
	\[             \bm p_{(\xi^{(r)})}^{(r)}=\frac{1}{\alpha^{(r)}}\left(  \A_{(\xi^{(r)})}^{(r+1)} - \A_{(\xi^{(r)})}^{(r)}  \right)  .          \]
	The inequality in \eqref{eq:descent} can be further written as
	\begin{align}\label{eq:descent_2}
	&\Phi(\bm \theta^{(r+1)})-\Phi(\bm \theta^{(r)})  \nonumber\\
	& \leq -\alpha^{(r)} \left\langle \nabla_{\A_{(\xi^{(r)})}} f(\bm \theta^{(r)})-\G_{(\xi^{(r)})}^{(r)},\bm p_{(\xi^{(r)})}^{(r)}-\bm P_{(\xi^{(r)})}^{(r)}\right\rangle\nonumber\\
	&\quad - \alpha^{(r)} \left\langle \nabla_{\A_{(\xi^{(r)})}} f(\bm \theta^{(r)})-\G_{(\xi^{(r)})}^{(r)},\bm P_{(\xi^{(r)})}^{(r)}\right\rangle \nonumber\\
	&\quad + \left(\frac{L(\alpha^{(r)})^2}{2} - \frac{\alpha^{(r)}}{2}\right)\|\bm p_{(\xi^{(r)})}^{(r)}\|^2.
	\end{align}
	Again, taking expectation conditioning on the filtration ${\cal B}^{(r)}$ and $\xi^{(r)}$, 
	we can upper bound ${\textstyle \frac{1}{\alpha^{(r)} }} \big( \mathbb{E}_{\bm\zeta^{(r)}}\left[\Phi(\bm \theta^{(r+1)})|{\cal B}^{(r)},\xi^{(r)}\right]-\Phi(\bm \theta^{(r)}) \big)$ by
	\begin{align}\label{eq:descent_3}
	& \mathbb{E}_{\bm\zeta^{(r)}}\left[\left\langle \nabla_{\A_{(\xi^{(r)})}} f(\bm \theta^{(r)})-\G_{(\xi^{(r)})}^{(r)},\bm P_{(\xi^{(r)})}^{(r)}-\bm p_{(\xi^{(r)})}^{(r)}\right\rangle|{\cal B}^{(r)},\xi^{(r)}\right]\nonumber\\
	&\quad + \left(\frac{L \alpha^{(r)}}{2} - \frac{1}{2}\right)\mathbb{E}_{\bm\zeta^{(r)}}\left[\left\|\bm p_{(\xi^{(r)})}^{(r)}\right\|^2\left|{\cal B}^{(r)},\xi^{(r)}\right.\right],
	\end{align}
	i.e., the second term on the right hand side of \eqref{eq:descent_2} becomes zero because of Fact~\ref{fact:unbias}.
	The first term of \eqref{eq:descent_3} can be bounded via the following chain of inequalities:
	\begin{align}
	&\mathbb{E}_{\bm\zeta^{(r)}}\left[\left\langle \nabla_{\A_{(\xi^{(r)})}} f(\bm \theta^{(r)})-\G_{(\xi^{(r)})}^{(r)},\bm P_{(\xi^{(r)})}^{(r)}-\bm p_{(\xi^{(r)})}^{(r)}\right\rangle \left|{\cal B}^{(r)},\xi^{(r)} \right.\right] \nonumber \\
	&\quad \overset{(a)}{\leq} \mathbb{E}_{\bm\zeta^{(r)}}\left[ \left\|\bm \delta^{(r)} \right\|\left\|\bm P_{(\xi^{(r)})}^{(r)}-\bm p_{(\xi^{(r)})}^{(r)}\right\| ~\left|~{\cal B}^{(r)},\xi^{(r)}\right.\right] \nonumber\\
	&\quad \overset{(b)}{\leq}  \mathbb{E}_{\bm\zeta^{(r)}}\left[ \|\bm \delta^{(r)}\|^2 ~|~{\cal B}^{(r)},\xi^{(r)}\right] \leq (\sigma^{(r)})^2 
	\end{align}
	where (a) is due to the Cauchy-Schwartz inequality,
	and (b) is a consequence of the non-expansiveness of the proximal operator of convex $h_n(\cdot)$.
	Taking the total expectation, we have
	\begin{align}\label{eq:descent_4}
	&\mathbb{E}\left[\Phi(\bm \theta^{(r+1)})\right]-\mathbb{E}\left[ \Phi(\bm \theta^{(r)}) \right]  \\
	& \leq \alpha^{(r)}(\sigma^{(r)})^2 + \left(\frac{L(\alpha^{(r)})^2}{2} - \frac{\alpha^{(r)}}{2}\right)\mathbb{E} \left[\left\|\bm p_{(\xi^{(r)})}^{(r)}\right\|^2\right].\nonumber
	\end{align}

	Summing up the inequality from $t=0$ to $t=r-1$, 
	\begin{align}\label{eq:descent_5}
	&\mathbb{E}\left[\Phi(\bm \theta^{(r)})\right]- \Phi(\bm \theta^{(0)})  \\
	& \leq \sum_{t=0}^r\alpha^{(t)}(\sigma^{(t)})^2 + \sum_{t=0}^r \left(\frac{L(\alpha^{(t)})^2}{2} - \frac{\alpha^{(t)} }{2}\right)\mathbb{E}\left[\left\|\bm p_{(\xi^{(t)})}^{(t)}\right\|^2\right].\nonumber
	\end{align}
	Since $\alpha^{(r)}< 1/L$, we have $\frac{L(\alpha^{(r)})^2}{2} - \frac{\alpha^{(r)} }{2}<0$, therefore, 
	\begin{align}
	&\sum_{t=0}^r \left(  \frac{\alpha^{(t)} }{2}- \frac{L(\alpha^{(t)})^2}{2}\right)\mathbb{E}\left[\left\|\bm p_{(\xi^{(t)})}^{(t)}\right\|^2\right]\nonumber\\
	&\leq \Phi(\bm \theta^{(0)}) -\Phi(\bm \theta^{\star}) +\sum_{t=0}^r\alpha^{(t)}(\sigma^{(t)})^2,
	\end{align}
	such that $\bm \theta^{\star} \in \arg \min_{ \bm{\theta} } \Phi( \bm{\theta} )$.
	Taking $r\rightarrow \infty$, and
	by the assumption that $\sum_{r=0}^\infty \alpha^{(r)}(\sigma^{(r)})^2<\infty$, we can conclude that
	\[      \displaystyle \liminf_{r \rightarrow \infty} \mathbb{E}\left[\left\|\bm p_{(\xi^{(r)})}^{(r)}\right\|^2\right] =0,     \]
	using Lemma~\ref{lem:nnsequence}.
	
	To complete the proof, we observe that
	\begin{align}
	& \frac{1}{2} \mathbb{E}\left[\left\|\bm P_{(\xi^{(r)})}^{(r)}\right\|^2\right] \leq \mathbb{E}\left[\left\|\bm p_{(\xi^{(r)})}^{(r)}\right\|^2\right]+ \mathbb{E}\left[\left\|\bm p_{(\xi^{(r)})}^{(r)}-\bm P_{(\xi^{(r)})}^{(r)}\right\|^2\right] \nonumber \\
	&\leq \mathbb{E}\left[\left\|\bm p_{(\xi^{(r)})}^{(r)}\right\|^2\right]
	\nonumber\\&\quad + \mathbb{E}_{ \xi^{(r)}, {\cal B}^{(r)} } \left[\mathbb{E}_{\bm\zeta^{(r)}}\left[\left\|\bm G_{(\xi^{(r)})}^{(r)}-\nabla_{\A_{(\xi^{(r)})}}f(\bm \theta^{(r)})\right\|^2\left|{\cal B}^{(r)},\xi^{(r)}\right.\right]\right] \nonumber\\
	&\leq \mathbb{E}\left[\left\|\bm p_{(\xi^{(r)})}^{(r)}\right\|^2\right] + (\sigma^{(r)})^2.
	\end{align}
	where the last inequality is obtained via applying the nonexpansive property again.
	Note that both terms on the right hand side converge to zero.
	Hence, this relationship implies that
	\[     \displaystyle \liminf_{r \rightarrow \infty} \mathbb{E}\left[\left\|\bm P_{(\xi^{(r)})}^{(r)}\right\|^2\right] = 0.    \]
	Note that by our sampling strategy, we have
	\begin{align*}
	\mathbb{E}\left[\left\|\bm P_{(\xi^{(r)})}^{(r)}\right\|^2\right] & = \mathbb{E}_{\xi^{(r)}, {\cal B}^{(r)}} \left[ \mathbb{E}_{\bm\zeta^{(r)}} \left[\left\|\bm P_{(\xi^{(r)})}^{(r)}\right\|^2 \left|{\cal B}^{(r)},\xi^{(r)}\right.\right]\right].
	\end{align*}
	However, since $\bm P_{(\xi^{(r)})}^{(r)}$ is independent of the random seed $\zeta^{(r)}$, we have
	\begin{align*}
	\mathbb{E}\left[\left\|\bm P_{(\xi^{(r)})}^{(r)}\right\|^2\right] & = \mathbb{E}_{{\cal B}^{(r)}}\left[  \mathbb{E}_{\xi^{(r)}} \left[\left\|\bm P_{(\xi^{(r)})}^{(r)}\right\|^2 \left|{\cal B}^{(r)}\right.\right]\right]\nonumber\\
	& =  \mathbb{E}_{{\cal B}^{(r)}} \left[  \sum_{n=1}^N\frac{1}{N}\left\|\bm P_{(n)}^{(r)}\right\|^2\right].
	\end{align*}
	This proves the proposition.

	\section{Proof of Proposition~\ref{prop:adagrad}}\label{eq:adacpd}
	The insight of the proof largely follows the technique for single-block \texttt{Adagrad} \cite{li2018convergence}, with some careful modifications to multiple block updates.
	One will see that the block sampling strategy and the block-wise unbiased gradient estimation are key to apply the proof techniques developed in \cite{li2018convergence} to our case.
	To show convergence, let us first consider:
	\begin{Lemma}\cite{li2018convergence}\label{lem:ada_seq}
		Let $a_0>0$, $a_i\geq 0$, $i=1,\ldots,T$ and $\beta>1$. Then, we have
		\[       \sum_{t=1}^T\frac{a_t}{(a_0 + \sum_{i=1}^t a_i)^\beta}\leq \frac{1}{(\beta-1)a_0^{\beta-1}}.   \]
	\end{Lemma}
	The proof is simple and elegant; see \cite[Lemma 4]{li2018convergence}.

	\begin{Lemma}\label{lem:expectation}
		\cite{li2018convergence}
		Consider a random variable $X$.
		If $ \mathbb{E}[X]<\infty $, then $${\sf Pr}(X<\infty)=1.$$
	\end{Lemma}
	
	Let us consider the block-wise again:
	\begin{align}
	f({\bm \theta}^{(r+1)})&\leq f({\bm \theta}^{(r)}) +\left \langle \nabla_{\A_{(\xi^{(r)})}} f({\bm \theta}^{(r)}),\A_{(\xi^{(r)})}^{(r+1)}-\A_{(\xi^{(r)})}^{(r)} \right\rangle \nonumber\\
	& \quad\quad+ \frac{L_{\xi^{(r)}}^{(r)}}{2}\left\|\A_{(\xi^{(r)})}^{(r+1)}-\A_{(\xi^{(r)})}^{(r)}\right\|^2.
	\end{align}
	Plugging in our update rule under \texttt{AdaCPD},
	one can see that
	\begin{align}
	&f({\bm \theta}^{(r+1)})\leq f({\bm \theta}^{(r)}) +\left \langle \nabla_{\A_{(\xi^{(r)})}} f({\bm \theta}^{(r)}),-\bm \eta^{(r)}_{(\xi^{(r)})} \circledast \G_{(\xi^{(r)})}^{(r)} \right\rangle \nonumber\\
	& \quad\quad\quad\quad+ \frac{L_{\xi^{(r)}}^{(r)}}{2}\left\|\bm \eta^{(r)}_{(\xi^{(r)})} \circledast \G_{(\xi^{(r)})}^{(r)} \right\|^2 \nonumber\\
	& =  f({\bm \theta}^{(r)}) -\left \langle \nabla_{\A_{(\xi^{(r)})}} f({\bm \theta}^{(r)}),\bm \eta^{(r)}_{(\xi^{(r)})} \circledast\nabla_{\A_{(\xi^{(r)})}} f({\bm \theta}^{(r)})\right\rangle \nonumber\\
	&\quad +  \left \langle \nabla_{\A_{(\xi^{(r)})}} f({\bm \theta}^{(r)}),\bm \eta^{(r)}_{(\xi^{(r)})} \circledast \left(\nabla_{\A_{(\xi^{(r)})}} f({\bm \theta}^{(r)})-\G_{(\xi^{(r)})}^{(r)}\right) \right\rangle \nonumber\\
	& \quad + \frac{L_{\xi^{(r)}}^{(r)}}{2}\left\|\bm \eta^{(r)}_{(\xi^{(r)})} \circledast \G_{(\xi^{(r)})}^{(r)} \right\|^2.
	\end{align}

	Taking expectation w.r.t. $\bm\zeta^{(r)}$ (the random seed that is responsible for selecting fibers) conditioning on the filtration ${\cal B}^{(r)}$ and the selected block $\xi^{(r)}$, the middle term is zero---since the 
	block stochastic gradient is unbiased [cf. Fact~\ref{fact:unbias}].
	Hence, we have reached the following
	
	\begin{align}
	&\mathbb{E}_{\bm\zeta^{(r)}}\left[f({\bm \theta}^{(r+1)})|{\cal B}^{(r)},\xi^{(r)}\right]\leq   \mathbb{E}_{\bm\zeta^{(r)}}\left[f({\bm \theta}^{(r)})|{\cal B}^{(r)},\xi^{(r)}\right]\nonumber\\
	&  -\mathbb{E}_{\bm\zeta^{(r)}}\left[\left \langle \nabla_{\A_{(\xi^{(r)})}} f({\bm \theta}^{(r)}),\bm \eta^{(r)}_{(\xi^{(r)})} \circledast\nabla_{\A_{(\xi^{(r)})}} f({\bm \theta}^{(r)})\right\rangle |{\cal B}^{(r)},\xi^{(r)}\right] \nonumber\\
	& + \frac{L_{\xi^{(r)}}^{(r)}}{2}\mathbb{E}_{\bm\zeta^{(r)}}\left[\left\|\bm \eta^{(r)}_{(\xi^{(r)})} \circledast \G_{(\xi^{(r)})}^{(r)} \right\|^2 \left|{\cal B}^{(r)},\xi^{(r)}\right.\right].\label{eq:adagrad_descent}
	\end{align}
	Taking total expectation on both sides, we have
	\begin{align}
	&\mathbb{E}\left[f({\bm \theta}^{(r+1)})\right]\leq   \mathbb{E}\left[f({\bm \theta}^{(r)})\right]\nonumber\\
	&\quad\quad -\mathbb{E}\left[\left \langle \nabla_{\A_{(\xi^{(r)})}} f({\bm \theta}^{(r)}),\bm \eta^{(r)}_{(\xi^{(r)})} \circledast\nabla_{\A_{(\xi^{(r)})}} f({\bm \theta}^{(r)})\right\rangle \right] \nonumber\\
	& \quad + \mathbb{E}\left[\frac{L_{\xi^{(r)}}^{(r)}}{2}\left\|\bm \eta^{(r)}_{(\xi^{(r)})}\circledast \G_{(\xi^{(r)})}^{(r)} \right\|^2 \right].
	\end{align}
	From the above inequality and the assumption that $L^{(r)}_{(n)}$ is bounded from above by $L$, we can conclude that
	\begin{align*}
	&\sum_{r=0}^R \mathbb{E}\left[\left\langle\nabla_{\A_{(\xi^{(r)})}} f({\bm \theta}^{(r)}),\bm \eta^{(r)}_{(\xi^{(r)})} \circledast \nabla_{\A_{(\xi^{(r)})}} f({\bm \theta}^{(r)})\right\rangle \right]\nonumber\\
	&\leq f(\bm \theta^{(0)}) - f(\bm \theta^{(\star)})+  \sum_{r=0}^R\frac{L}{2}\mathbb{E}\left[\left\|\bm \eta^{(r)}_{(\xi^{(r)})} \circledast \G_{(\xi^{(r)})}^{(r)} \right\|^2 \right]
	\end{align*}
	by summing up all the inequalities in \eqref{eq:adagrad_descent} from $r=0$ to $R$.
	
	Taking $R\rightarrow \infty$ and observe that:
	\begin{align}
	&\mathbb{E}\left[ \sum_{r=0}^\infty\left\| \bm\eta^{(r)}_{(\xi^{(r)})} \circledast \G_{(\xi^{(r)})}^{(r)} \right\|^2 \right]\\
	& = \sum_{r=0}^\infty \mathbb{E} \left[ \left\| \left( \bm\eta^{(r+1)}_{(\xi^{(r)})} + \bm\eta^{(r)}_{(\xi^{(r)})} - \bm\eta^{(r+1)}_{(\xi^{(r)})} \right) \circledast \G_{(\xi^{(r)})}^{(r)} \right\|^2 \right] \nonumber \\
	& = \sum_{r=0}^\infty \mathbb{E} \left[ \left\| \bm\eta^{(r+1)}_{(\xi^{(r)})} \circledast \G_{(\xi^{(r)})}^{(r)} \right\|^2 \right] \nonumber \\
	& \quad \quad + \sum_{r=0}^\infty  \mathbb{E} \left[ \left\| \left( \bm\eta^{(r)}_{(\xi^{(r)})} - \bm\eta^{(r+1)}_{(\xi^{(r)})} \right) \circledast \G_{(\xi^{(r)})}^{(r)} \right\|^2 \right].
	\end{align}
	
	Note that we have exchanged the order of the limits and expectations, since the expectation is taking on nonnegative terms.
	Using Lemma~\ref{lem:ada_seq}, one can easily show the first term above is bounded from above by
	$ \frac{C_1}{2\epsilon \beta^{2\epsilon}}$,
	where $0<C_1<\infty$ is a constant.
	To see the second term is bounded, observe 
	\begin{align}
	&\mathbb{E}\left[ \sum_{r=0}^\infty\sum_{i=1}^{J_{\xi^{(r)}}}\sum_{f=1}^F\left( \left[\bm \eta^{(r)}_{(\xi^{(r)})}\right]_{i,f}^2-\left[\bm \eta^{(r+1)}_{(\xi^{(r)})}\right]_{i,f}^2\right)\left[\G_{(\xi^{(r)})}^{(r)}\right]_{i,f}^2\right ] \nonumber\\
	&= \widetilde{\mathbb{E}}\left[\sum_{r=0}^\infty \sum_{f=1}^F \frac{1}{N}\sum_{n=1}^N \sum_{i=1}^{J_n} \left( \left[\bm \eta^{(r)}_{(n)}\right]_{i,f}^2-\left[\bm \eta^{(r+1)}_{(n)}\right]_{i,f}^2\right)\left[\G_{(n)}^{(r)}\right]_{i,f}^2 \right ] \nonumber\\
	&\leq  \widetilde{\mathbb{E}}\left[\sum_{n=1}^N \sum_{i=1}^{J_n}\sum_{f=1}^F \frac{1}{N}\max_{r\geq 0}  \left[\G_{(n)}^{(r)}\right]_{i,f}^2 \sum_{r=0}^\infty \left( \left[\bm \eta^{(r)}_{(n)}\right]_{i,f}^2-\left[\bm \eta^{(r+1)}_{(n)}\right]_{i,f}^2\right)\right] \nonumber \\
	&\leq \widetilde{\mathbb{E}}\left[\sum_{n=1}^N \sum_{i=1}^{J_n}\sum_{f=1}^F \frac{1}{N}\max_{r\geq 0}  \left[\G_{(n)}^{(r)}\right]_{i,f}^2 \left[\bm \eta^{(0)}_{(n)}\right]_{i,f}^2\right] \nonumber \\
	&\leq  \sum_{n=1}^N \sum_{i=1}^{J_n}\sum_{f=1}^F \frac{2}{N}\left[\bm \eta^{(0)}_{(n)}\right]_{i,f}^2\widetilde{\mathbb{E}}\left[\max_{r\geq 0}  \left[ \left[  \nabla_{\A_{(n)}} f(\bm \theta^{(r)})\right]_{i,f}^2 \right.\right.\nonumber \\
	&\quad \left. \left.+\left( \left[  \nabla_{\A_{(n)}} f(\bm \theta^{(r)})\right]_{i,f}-\left[\G_{(n)}^{(r)}\right]_{i,f}\right)^2 \right]\right] \nonumber,	\end{align}
	where $\widetilde{\mathbb{E}}$ means taking expectation w.r.t. all the random variables except for $\xi^{(r)}$ for $r=0,\ldots,\infty$,
	and the second inequality is due to the effect of the telescope summation.
	Since we have assumed that $\A_{(n)}^{(r)}$'s are bounded, the right hand side is bounded from above.
	Therefore, we have reached the conclusion 
	\[ \widetilde{\mathbb{E}}\left[\sum_{r=0}^\infty \left\langle\nabla_{\A_{(\xi^{(r)})}} f({\bm \theta}^{(r)}),\bm \eta^{(r)}_{(\xi^{(r)})} \circledast \nabla_{\A_{(\xi^{(r)})}} f({\bm \theta}^{(r)})\right\rangle \right] <\infty. \]
	
	Applying Lemma~\ref{lem:expectation}, one can see that
	\[      {\sf Pr}\left(   \sum_{r=0}^\infty    \left[ \bm \eta^{(r)}_{(\xi^{(r)})} \right]_{i,f}\left[ \nabla_{\A_{(\xi^{(r)})}} f({\bm \theta}^{(r)})\right]_{i,f}^2  <\infty      \right)  =1.     \]
	Since ${\sf Pr}(\xi^{(r)}=n)>0$, one immediate result is that any $n$ appears infinitely many times in the sequence $r=0,\ldots,\infty$, according to the second Borel-Cantelli lemma.
	This leads to 
	\[      {\sf Pr}\left(   \sum_{j=1}^\infty    \left[ \bm \eta^{(r_j(n))}_{(n)} \right]_{i,f}\left[ \nabla_{\A_{(n)}} f({\bm \theta}^{(r_j(n))})\right]_{i,f}^2  <\infty      \right)  =1,     \]
	holds for $n=1,\ldots,N$,
	where $r_1(n),\ldots,r_j(n),\ldots$ is the subsequence of $\{r\}$ such that block $n$ is sampled for updating.
	Hence, with probability one there exists a subsequence $r_1(n),\ldots,r_\infty(n)$ such that at the corresponding iterations block $n$ is sampled for updating.
	It is not hard to show that
	\[  \sum_{j=1}^\infty   \left[ \bm \eta^{(r_j(n))}_{(n)} \right]_{i,f} =\infty,  \]
	by the assumption that $\A_{(n)}^{(r)}$ are all bounded.
	This directly implies that $\sum_{r=1}^\infty   \left[ \bm \eta^{(r)}_{(n)} \right]_{i,f} =\infty$. Together with Lemma~\ref{lem:nnsequence}, we have
	\[ {\sf Pr} \left( \liminf_{r \rightarrow \infty}~[\nabla_{\A_{(n)}} f({\bm \theta}^{(r)})]_{i,f}^2= 0 \right) = 1,~~\forall~i,f. \]
	
	\bibliographystyle{IEEEtran}

	\vspace{-1.3cm}
	\begin{IEEEbiography}[{\includegraphics[width=1in,height=1.25in,clip,keepaspectratio]{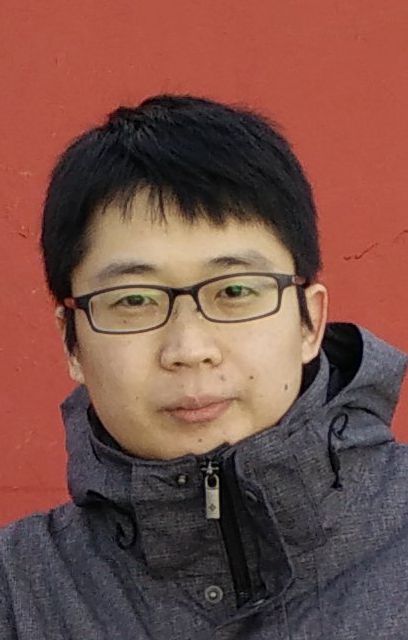}}]
	{Xiao Fu} (S'12-M'15) is an Assistant Professor in the School of Electrical Engineering and Computer Science, Oregon State University, Corvallis, Oregon, United States. He received his Ph.D. degree in Electronic Engineering from The Chinese University of Hong Kong (CUHK), Hong Kong, in 2014. He was a Postdoctoral Associate in the Department of Electrical and Computer Engineering, University of Minnesota, Minneapolis, MN, United States, from 2014 to 2017. His research interests include the broad area of signal processing and machine learning. He received a Best Student Paper Award at ICASSP 2014. Two of his co-authored papers received Best Student Paper 
	Awards at IEEE CAMSAP 2015 and IEEE MLSP 2019, respectively. He serves as the treasurer of IEEE Signal Processing Society Oregon Chapter. He also serves as a member of the EURASIP Technical Area Committee in Signal Processing for Multisensor Systems for the term of 2020-2023. He is a member of IEEE.
\end{IEEEbiography}
\vspace{-0.9cm}
	\begin{IEEEbiography}[{\includegraphics[width=1in,height=1.25in,clip,keepaspectratio]{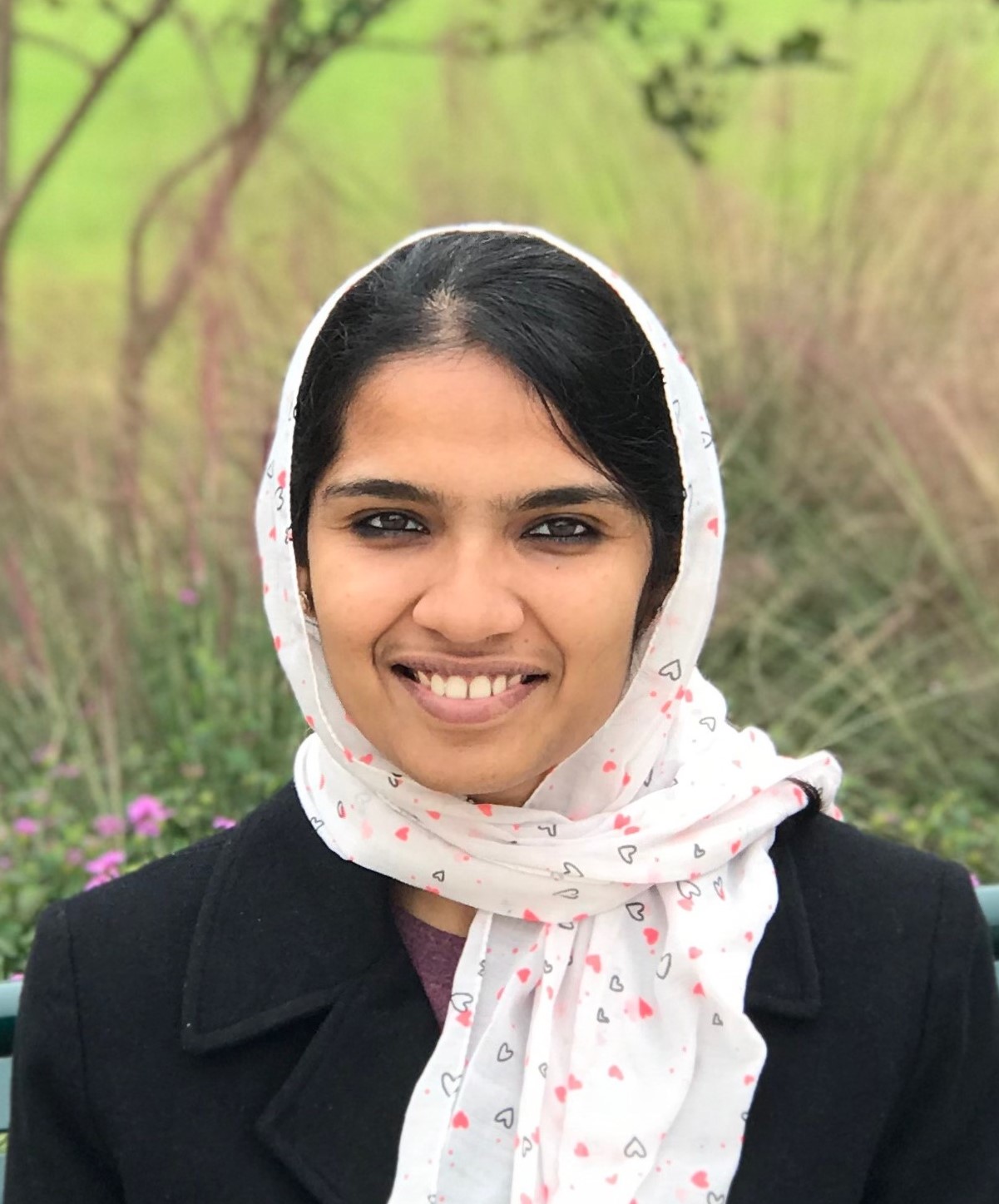}}]
	{Shahana Ibrahim} received her B.Tech. degree in Electronics and Communication Engineering from National Institute of Technology, Calicut, India, in 2012. She had been working as System Validation Engineer at Texas Instruments, Bengaluru, India, from 2012 to 2017. She received her M.S. degree in Electrical Engineering from Oregon State University, Corvallis, Oregon, United States, in 2019. She is currently pursuing her PhD degree in Electrical Engineering at Oregon State University, Corvallis, Oregon, United States. Her research interests are in the broad areas of statistical machine learning and signal processing.
\end{IEEEbiography}
\vspace{-0.9cm}
	\begin{IEEEbiography}[{\includegraphics[width=1in,height=1.20in,clip,keepaspectratio]{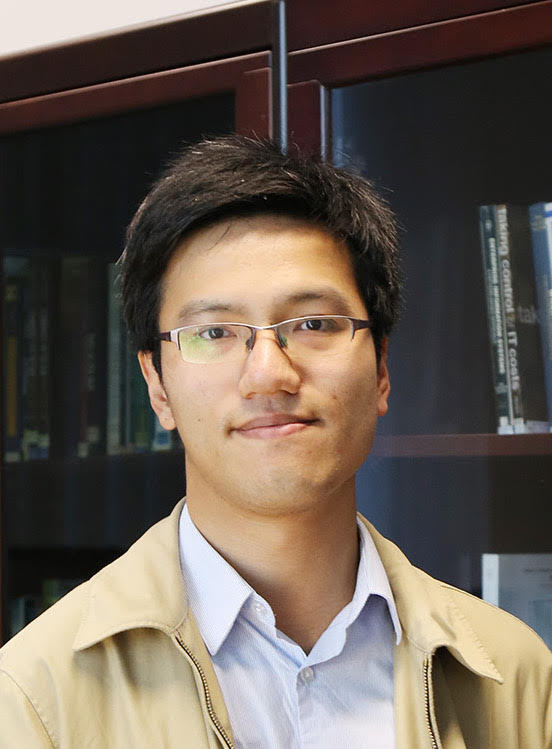}}]
	{Hoi-To Wai} (S’11–M’18) received his PhD degree from Arizona State University (ASU) in Electrical Engineering in Fall 2017, B. Eng. (with First Class Honor) and M. Phil. degrees in Electronic Engineering from The Chinese University of Hong Kong (CUHK) in 2010 and 2012, respectively. He is an Assistant Professor in the Department of Systems Engineering \& Engineering Management at CUHK. He has held research positions at ASU, UC Davis, Telecom ParisTech, Ecole Polytechnique, LIDS, MIT.
Hoi-To's research interests are in the broad area of signal processing, machine learning and distributed optimization, with a focus of their applications to network science. His dissertation has received the 2017's Dean's Dissertation Award from the Ira A. Fulton Schools of Engineering of ASU and he is a recipient of a Best Student Paper Award at ICASSP 2018.\vspace{-0cm}
\end{IEEEbiography}
\vspace{-1.1cm}
	\begin{IEEEbiography}[{\includegraphics[width=1in,height=1.25in,clip,keepaspectratio]{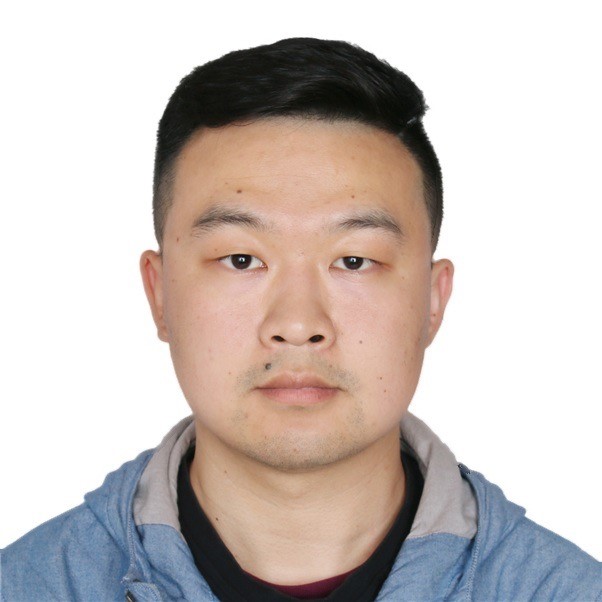}}]
	{Cheng Gao} is a Computer Science PhD student at University of Missouri, Columbia, Missouri, United States. He is currently focusing his research on sparse learning and bioinformatics. 
He earned his Master's degree in Electrical and Computer Engineering from Oregon State University, Corvallis, Oregon, United States, in 2019 with research focus on large-scale tensor decomposition algorithms. He received his Bachelor’s degree in Automation from Wuhan University of Technology, China in 2016.
\end{IEEEbiography}
\vspace{-1.1cm}
	\begin{IEEEbiography}[{\includegraphics[width=1in,height=1.25in,clip,keepaspectratio]{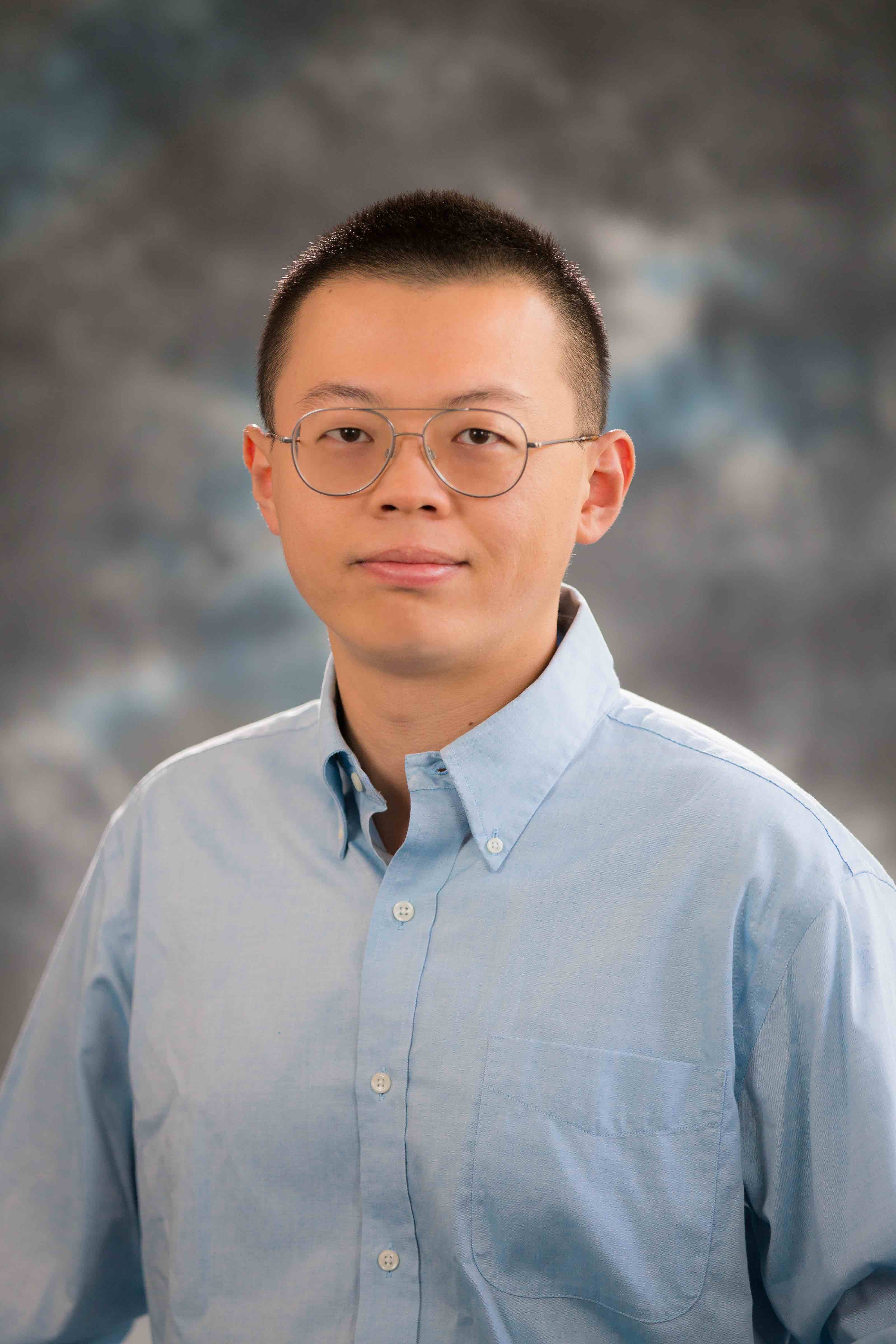}}]
	{Kejun Huang} received his B.Eng. degree in communication engineering from the Nanjing University of Information Science and Technology, China, in 2010 and his Ph.D. degree in electrical engineering from the University of Minnesota, Minneapolis, in 2016. He is an assistant professor in the Department of Computer and Information Science and Engineering at the University of Florida, Gainesville. He was a postdoctoral associate in the Department of Electrical and Computer Engineering at the University of Minnesota, Minneapolis, from 2016 to 2018. His research interests include machine learning, signal processing, optimization, and statistics. He is a Member of the IEEE.
\end{IEEEbiography}
\end{document}